\documentclass[aps,pre,twocolumn,superscriptaddress,showkeys,amsmath,amssymb]{revtex4}
\usepackage[unicode,pdfstartview=FitH,bookmarksnumbered,breaklinks,colorlinks,linktocpage,citecolor=blue,linkcolor=red,urlcolor=blue]{hyperref}

\usepackage{epsfig,amsmath,amssymb,color}
\usepackage[T1]{fontenc}
\usepackage[latin9]{inputenc}

\begin{document}

\title{Towards a quasiphase transition in the single-file chain of water molecules:\\
       Simple lattice model}

\author{Maksym~Druchok}
\email{maksym@icmp.lviv.ua}
\affiliation{Institute for Condensed Matter Physics,
         National Academy of Sciences of Ukraine,
         Svientsitskii Street 1, 79011 L'viv, Ukraine}

\author{Volodymyr~Krasnov}
\affiliation{Institute for Condensed Matter Physics,
          National Academy of Sciences of Ukraine,
          Svientsitskii Street 1, 79011 L'viv, Ukraine}

\author{Taras~Krokhmalskii}
\affiliation{Institute for Condensed Matter Physics,
          National Academy of Sciences of Ukraine,
          Svientsitskii Street 1, 79011 L'viv, Ukraine}

\author{Tatiana~Cardoso~e~Bufalo}
\affiliation{Departamento de Fisica,
          Universidade Federal de Lavras,
          CP 3037, 37200-000, Lavras-MG, Brazil}

\author{Sergio~Martins~de~Souza}
\affiliation{Departamento de Fisica,
          Universidade Federal de Lavras,
          CP 3037, 37200-000, Lavras-MG, Brazil}

\author{Onofre~Rojas}
\email{ors@ufla.br}
\affiliation{Departamento de Fisica,
          Universidade Federal de Lavras,
          CP 3037, 37200-000, Lavras-MG, Brazil}

\author{Oleg~Derzhko}
\email{derzhko@icmp.lviv.ua}
\affiliation{Institute for Condensed Matter Physics,
          National Academy of Sciences of Ukraine,
          Svientsitskii Street 1, 79011 L'viv, Ukraine}
\affiliation{Department of Metal Physics,
          Ivan Franko National University of L'viv,
          Kyrylo \& Mephodiy Street 8, 79005 L'viv, Ukraine}

\date{\today}

\begin{abstract}
Recently,
X.~Ma {\it et al.} [Phys. Rev. Lett. {\bf 118}, 027402 (2017)] have suggested
that water molecules encapsulated in (6,5) single-wall carbon nanotube experience a temperature-induced quasiphase transition around 150~K
interpreted as changes in the water dipoles orientation.
We discuss further this temperature-driven quasiphase transition
performing quantum chemical calculations and molecular dynamics simulations
and, most importantly, suggesting a simple lattice model to reproduce the properties of the one-dimensionally confined finite arrays of water molecules.
The lattice model takes into account not only the short-range and long-range interactions but also the rotations in a narrow tube
and the both ingredients provide an explanation for a temperature-driven orientational ordering of the water molecules,
which persists within a relatively wide temperature range.
\end{abstract}

\keywords{single-walled carbon nanotubes, single-file water molecules, orientational order, quasiphase transition}

\maketitle

\section{Introduction}
\label{sec1}

Confinement of water into narrow nanopores changes essentially its properties \cite{Wang2021}.
Water molecules may be confined to pores of nanometer diameters and form single-file chains.
Carbon nanotubes with nanometer-scale diameters provide an excellent experimental setup to study one-dimensionally confined water.
Recently,
it was demonstrated experimentally that nanometer diameters single-walled carbon nanotubes can be filled with water \cite{Cambre2010}.
Most intriguing is the behavior of water in single-walled carbon nanotubes with such diameters (around 0.5~nm)
for which mutual passage of water molecules is excluded.
Furthermore,
the electronic structure of single-walled carbon nanotubes (semiconducting or metallic) depends on their chirality.
Therefore, only due to recent advances in sorting of carbon nanotubes,
which provide empty and water-filled single-chirality single-walled carbon nanotubes with a well-defined small diameter,
a precise experimental study of the one-di\-men\-sional water becomes feasible.

A short while ago,
X.~Ma {\it et al.} \cite{Ma2017} have reported temperature-dependent photoluminescence spectroscopy data
for single-chirality (6,5) single-walled carbon nanotubes (CNTs).
Empty CNTs exhibit a linear tempe\-rature-dependent photoluminescence spectral shift as expected.
Water-filled CNTs show a stepwise photoluminescence spectral shift centered at about $150$~K
which is superimposed on the anticipated linear temperature-dependent one.
X.~Ma {\it et al.} \cite{Ma2017} assumed that the origin of the observed additional spectral shift
is related to a significant change in the orientation of the water dipoles.
Furthermore,
they performed molecular dynamics (MD) simulations to support the interpretation of the measured photoluminescence spectra at different temperatures.
Their MD calculations indicate three different regimes:
1) traditional hydrogen-bonded chains
(below $\sim 40$~K),
2) predominantly bifurcated hydrogen bonds
where hydrogen bond from a single oxygen atom is distributed over both hydrogen atoms of a single neighboring water molecule
(around $\sim 70$~K),
and
3) disordered chains
(for $T>200$~K).
The direction of the measured photoluminescence spectral shift
agrees with the effective total dipole moment of the structures dominated as temperature grows,
however,
more extended calculations are desired to shed further light on the nature of these quasiphase transitions.
We mention here a very recent attempt to examine ordering effects in a chain of $N=11$ water molecules
within a path integral ground state approach \cite{Sahoo2021}.
Furthermore, density matrix renormalization group calculations for longer chains are reported in Ref.~\cite{Serwatka2022a},
whereas quantum phases in the one-dimensional water chain are discussed in Refs.~\cite{Serwatka2022b,Serwatka2023}.

The aim of the present paper is to report various theoretical calculations
of the orientational ordering in a single-file of water molecules in (6,5) CNT.
To this end,
we first reconsider and extend the quantum chemistry and MD studies of Ref.~\cite{Ma2017} providing more details about these calculations.
Then, we introduce a lattice model for confined water quasiphase transitions in CNT,
which is able to mimic the temperature behavior of the orientational order of the water dipoles.
It should be mentioned here
that one-dimensional lattice models with point dipoles sitting on the lattice sites
(one-dimensional dipole models)
were introduced by J.~K\"{o}finger {\it et al.} and successfully used to examine
the filling-emptying transition,
bistability of the particle-number distribution,
static dielectric response to an external field \cite{Koefinger2008,Koefinger2009a,Koefinger2009b,Koefinger2010a,Koefinger2010b,Koefinger2011}.
However, those models capture the orientational defects in a hydrogen-bonded chain
and do not represent the configurations relevant for the orientational order around 150~K suggested in Ref.~\cite{Ma2017}.

The rest of the paper is organized as follows.
In Section~\ref{sec2},
we briefly describe quantum-chemical computations and MD simulations.
In Section~\ref{sec3},
we introduce a lattice model and discuss its various properties mainly from the perspective of a temperature-driven dipole ordering.
We conclude with a brief summary in Section~\ref{sec4}.

\section{Quantum chemistry and molecular dynamics calculations}
\label{sec2}

\subsection{Quantum chemistry}
\label{sec21}

To examine the water molecules encapsulated in (6,5) CNT by MD simulations,
the realistic charges for the water hydrogen and oxygen atoms are primarily required \cite{Ma2017}.
These charges can be obtained
either from relatively simple semi-empirical calculations,
from more complicate first-principle calculations,
or from even more demanding density functional theory.
Let us discuss this issue in some detail.

To obtain the charges for the hydrogen and oxygen atoms in water in the (6,5) CNT,
we performed the semi-empirical calculations using the GAMESS package \cite{Schmidt1993} and AM1 (Austin Model 1) method.
To model the (6,5) CNT,
we used the structure with 362 carbon atoms and 20 hydrogen atoms added to saturate free carbon bonds on the edges of CNT.
The coordinates of carbon atoms were frozen while for the coordinates of hydrogen atoms the optimal positions were found.
Then, the calculations for optimal positions of 11 water molecules inside the described above CNT were performed.
To avoid surface effects,
i.e., to minimize effects of molecules at terminal positions,
only charges of 7 inner water molecules were taken into account.
The average charges obtained are: $q_{\rm O}=-0.4348 e$ for oxygen atoms and $q_{\rm H}=-q_{\rm O}/2=0.2174 e$ for hydrogens
(here $e$ is the elementary electric charge).
It should be stressed that the values of $q_{\rm O}$ and $q_{\rm H}$ are significantly smaller than those values used in,
e.g.,
more common TIP3P water ($q_{\rm O}=-0.834 e$, $q_{\rm H}=0.417 e$) \cite{Jorgensen1983}
or
SPC/E water ($q_{\rm O}=-0.8476 e$, $q_{\rm H}=0.4238 e$).
Smaller charges result in a reduced dipole moment of the water molecule.
Since MD simulations,
except $q_{\rm O}$ and $q_{\rm H}$,
use other characteristics of all atoms constituting the water in CNT, which are already intrinsically optimized,
we simply augment the obtained charges by the SPC/E water geometry,
i.e., $\alpha_{\rm H-O-H}=109.47^{\circ}$ and $r_{\rm O-H}=1.00$~\AA,
getting for the value of the dipole moment $\mu=1.105$~D.
Recall, the dipole moment of the SPC/E water molecule is $\mu=2.351$~D.
Significantly smaller dipole moment of the water molecule in water-filled nanotubes in comparison to water wires without nanotubes
was reported in {\it ab initio} MD simulations of a water-filled (6,6) CNT in Ref.~\cite{Mann2003}
(see also Ref.~\cite{Moulin2005}).

To check what happens beyond the AM1 method,
we performed more quantum chemistry calculations
(semi-empirical, Hartree-Fock, and also density-functional-theory ones)
using the GAMESS package \cite{Schmidt1993} (see Supplementary Material).
The outcomes are rather diverse.
For instance,
the average charge for the oxygen atoms $q_{\rm O}$ varies from about $-0.2e$ to about $-0.9e$
depending on the specific procedure utilized.
The partial charges $q_{\rm O}$ and $q_{\rm H}$,
which are required as parameters for further simulations,
appear as a result of dividing up the overall molecular charge density into atomic contributions.
They strongly depend on
the choice of the basis set,
the quantum mechanical method,
the population analysis method,
as well as on the choice of the geometry of water molecule.
The charge distribution in the water molecule
(in gas phase, not in a CNT)
was analyzed in great detail in Ref.~\cite{Martin2005}.
On the other hand,
as it follows from MD simulations, see Sec.~\ref{sec22} below,
these charges are extremely important for very existence of the orientational order of water dipoles at intermediate temperatures.
Hence,
the precise determination of $q_{\rm O}$ and $q_{\rm H}$ as well as other characteristics of water molecules in CNT
remains an important issue to be resolved in the future.

\subsection{Molecular dynamics}
\label{sec22}

Our MD calculations reported below
not only reproduce the basic results of X.~Ma {\it et al.} \cite{Ma2017} with more details
but also provide other quantities illustrating behavior of water molecules confined to a single file inside CNTs.

\begin{figure}[htb!]
\begin{center}
\includegraphics[clip=true,angle=0,width=0.7\columnwidth]{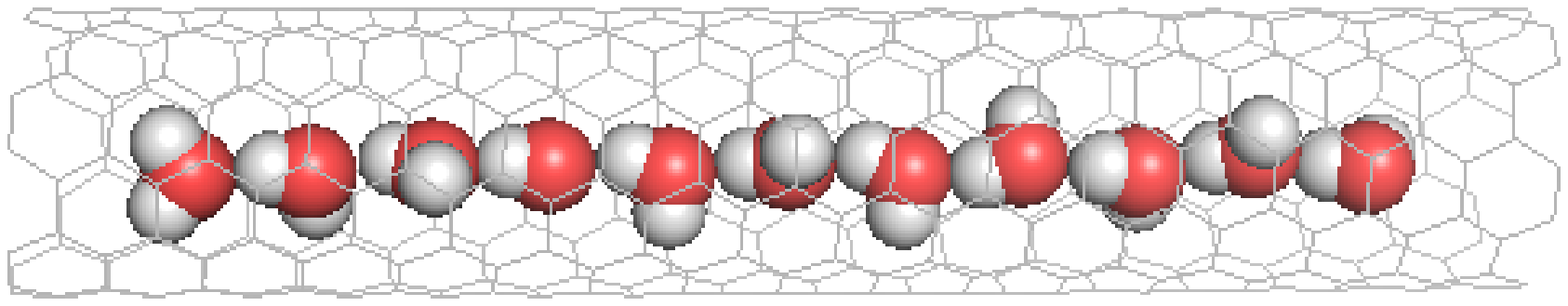}
\includegraphics[clip=true,angle=0,width=0.2\columnwidth]{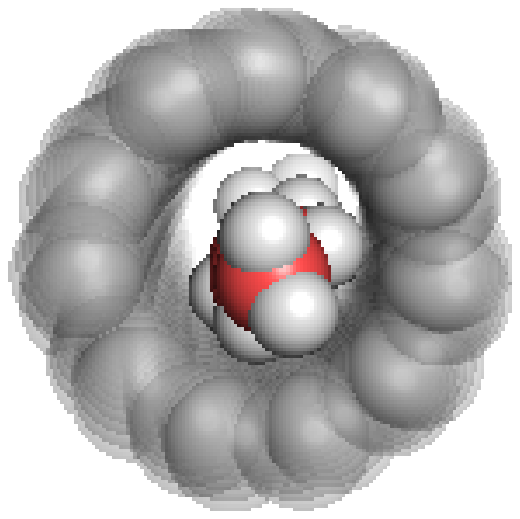}
\caption{Snapshot of $N=11$ water molecules encapsulated in (6,5) CNT of $\approx 40$~{\AA} at $T=10$~K.
Left and right views show CNT in different orientations and representations (wireframe versus spheres).
The right view is intended to showcase that the majority of inner space is occupied by water.
CNT sidewalls are depicted as wireframe (left view) or semitransparent spheres (right view), water molecules -- as spheres.
Hydrogen and oxygen atoms are shown in white and red, respectively.}
\label{fig01}
\end{center}
\end{figure}

We performed a series of MD simulations of water molecules encapsulated inside carbon nanotubes with the chirality of (6,5) at different temperatures.
The (6,5) CNTs have a diameter of $\approx 7.4$~\AA.
The stated value denotes the diameter of the circle over the centers of carbon atoms composing the CNT openings.
The actual interior, available for water molecules, is smaller due to van der Waals sizes of carbons.
Such a small-sized nanopore allows only a single file arrangement of water molecules.
In addition to the temperature effect we also considered different lengths of water chains and nanotubes.
Three different cases were examined:
i) $N{=}11$ water molecules encapsulated inside a CNT of $\approx 40$~{\AA}
(a sample snapshot is shown in Fig.~\ref{fig01}),
ii) $N{=}20$ water molecules inside a CNT of $\approx 85$~{\AA},
and
iii) $N{=}35$ water molecules inside a CNT of $\approx 170$~{\AA}.
Such a choice is intended to test whether the effect of a collective arrangement of water molecules persists
for longer sequences with various volumes per water molecule.
Therefore, for each of the cases listed above we ran a set of simulations over a temperature range of $10\ldots 240$~K.
Note that around the lower temperature boundary quantum fluctuations (which may be larger than thermal ones) should be taken into account
(e.g., using path integrals \cite{Ceperley1995});
therefore, our MD simulation results (as well as those reported in Ref.~\cite{Ma2017}) should be considered with caution around 10~K.
We also notice that quantum effects are important for water molecules even at higher temperatures,
e.g., when diffusion of protons is considered \cite{Rossi2016}
(see also Ref.~\cite{Ceriotti2016}).

We generated starting configurations consisting of CNTs and water molecules.
The short-range interactions for water were taken from the SPC/E water model \cite{Berendsen1987} (see also Ref.~\cite{Liu2016}),
while the charges
($-0.4348e$ and $0.2174e$ for oxygens and hydrogens, respective\-ly)
were optimized within the AM1 level of approximation,
see Sec.~\ref{sec21}.
The CNT model was taken from Ref.~\cite{Huang2006}, namely, the Lennard-Jones parameters for carbons of nanotube sidewalls.
Usually, the CNT simulations also imply a set of spring bonds and angles preserving the CNT geometry,
however, since waters are in the spot of interest we froze the ideal CNT in vacuum to cut the computational costs.
One has to mention that the sidewall carbons are neutral.
The Lennard-Jones parameters for unlike sites were calculated using the geometric mean mixing rules for both sigma and epsilon values.
This combination of interaction parameters was successfully utilized
in our recent studies \cite{Druchok2017,Druchok2018,Druchok2019} of nanotubes interacting with
SPC/E water.

The CNT sidewalls are hydrophobic, so the water molecules prefer to group together inside the CNT rather than spread over the volume.
However, we placed two additional carbon atoms at the centers of CNT ends
to assure that water molecules stay inside the nanotube interior during the whole run of the simulation.

The simulation conditions were kept the same for all the runs except the temperature variation.
The temperature was controlled by means of the NVT Nose--Hoover thermostat.
Each simulation utilized the leapfrog integration algorithm with the time step of 0.001~ps,
covering 25~ps of equilibration and then 200~ps of a production run.
Smooth particle mesh Ewald technique was used to calculate the electrostatic terms,
while the short-range interactions were calculated with the cut-off distance of 15~{\AA}.

\begin{figure}
\begin{center}
\includegraphics[clip=true,width=0.65\columnwidth]{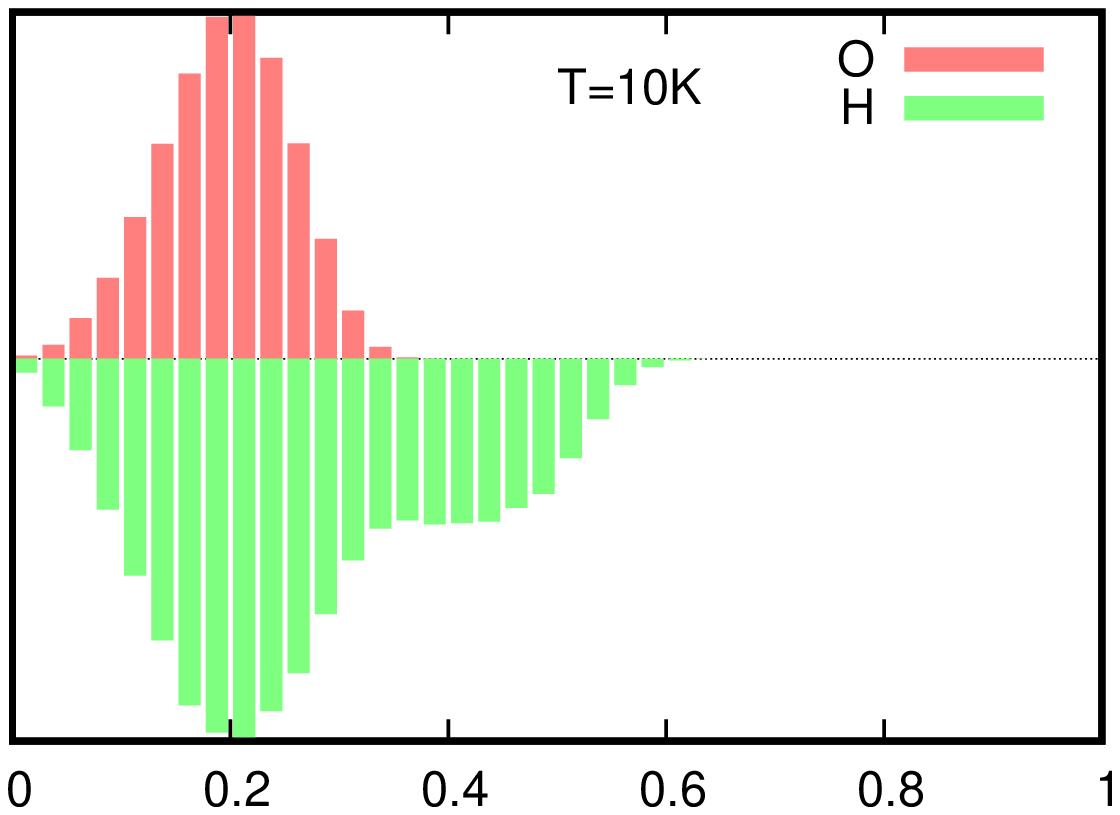}\\
\includegraphics[clip=true,width=0.65\columnwidth]{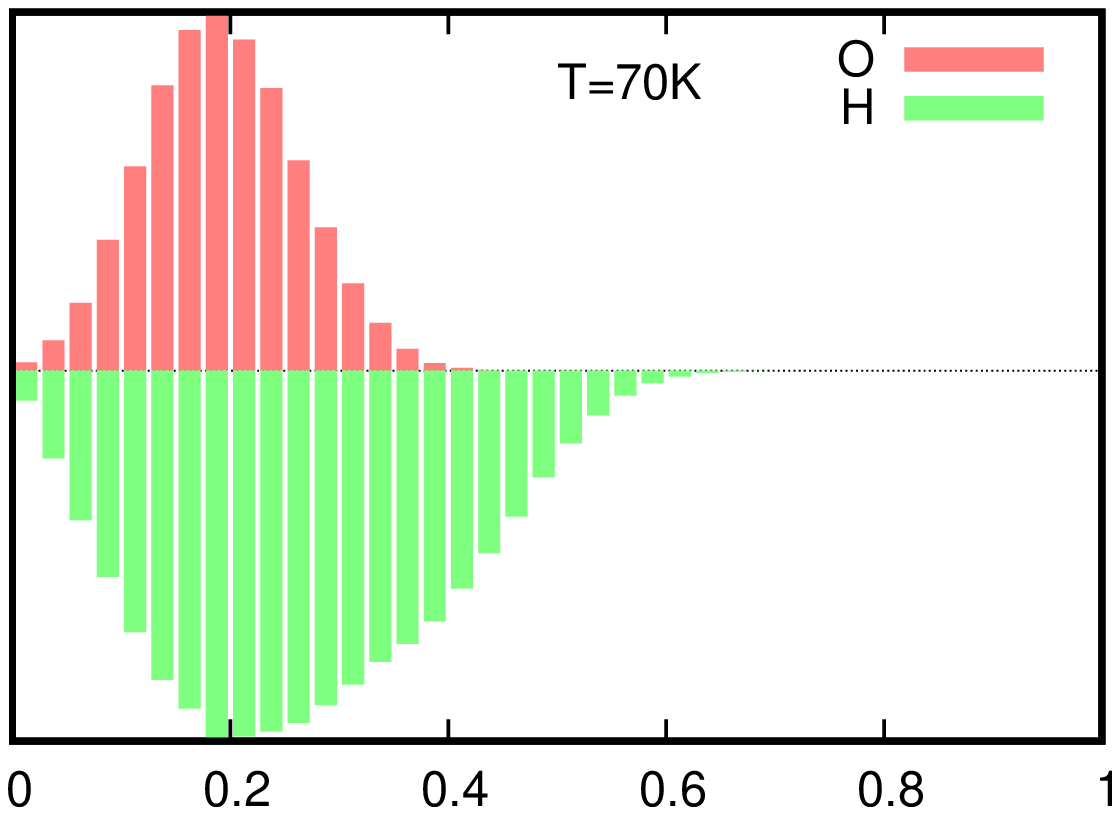}\\
\includegraphics[clip=true,width=0.65\columnwidth]{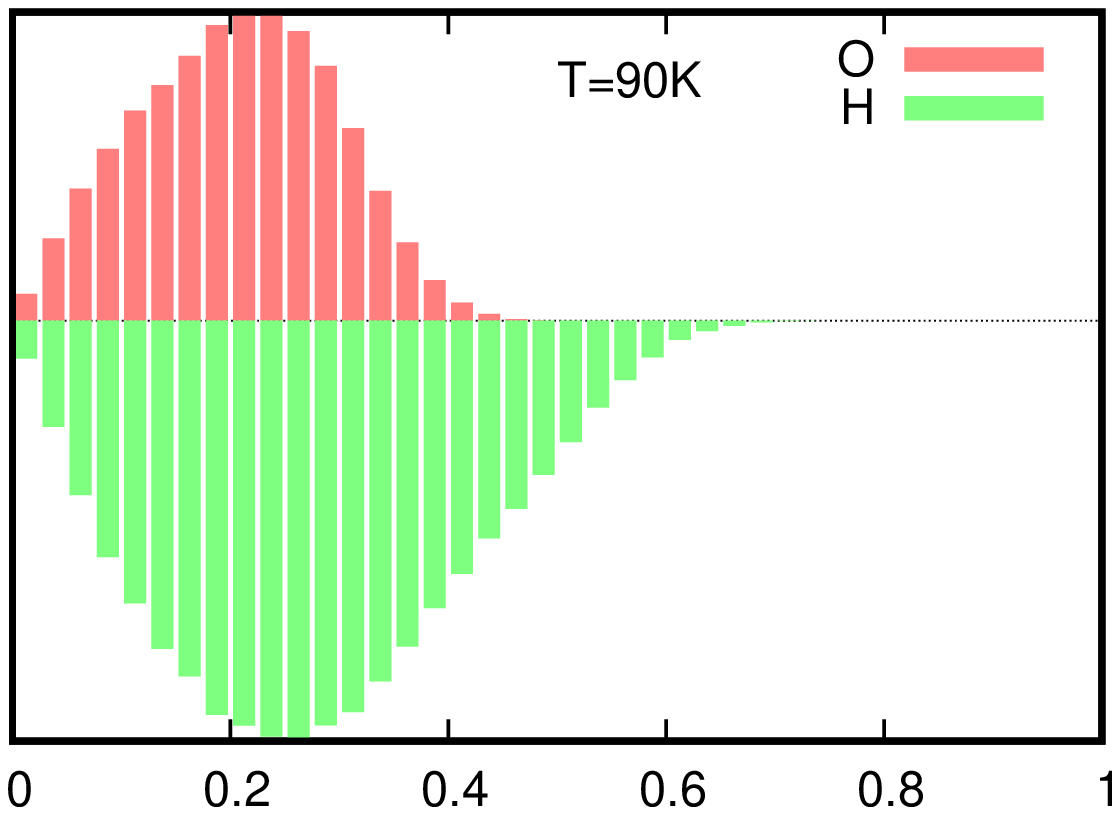}\\
\includegraphics[clip=true,width=0.65\columnwidth]{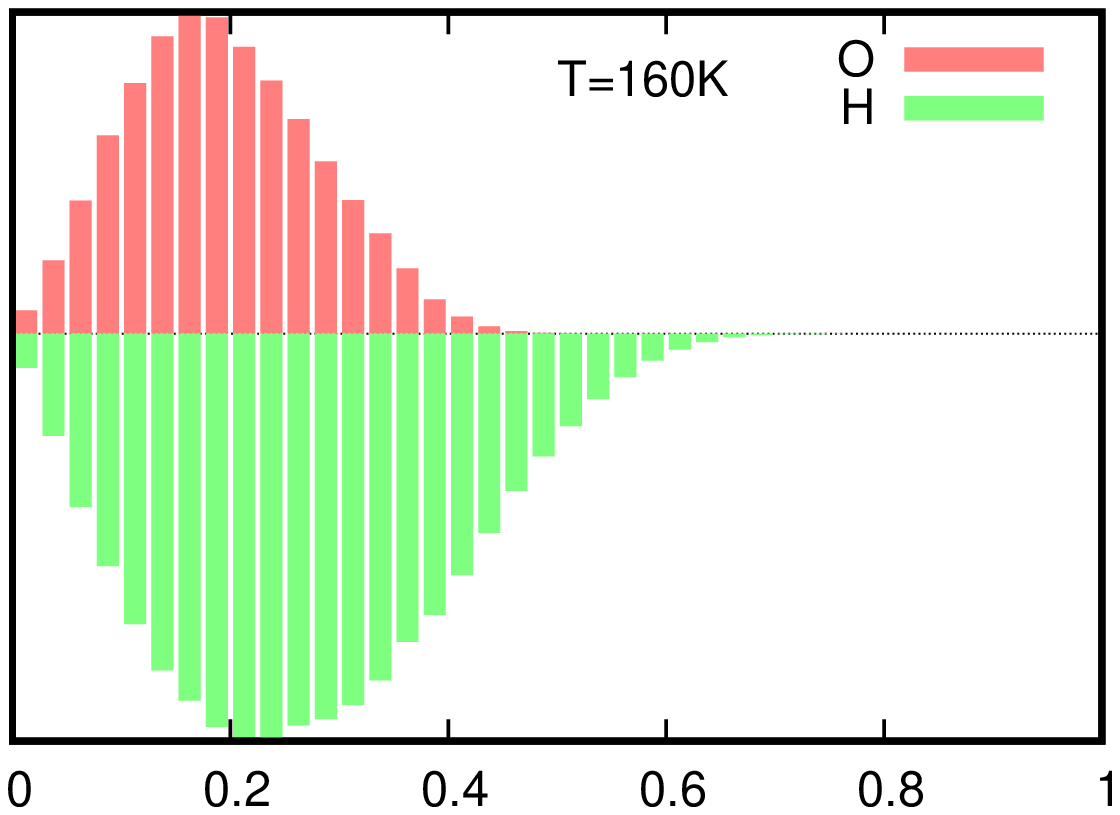}
\caption{The histograms showing preferential radial positions of oxygen (red) and hydrogen (green) atoms of water inside (6,5)~CNT.
The $x$ values are scaled to fit the range of [0:1], so $x=0$ corresponds to the CNT axis, $x=1$ -- the CNT sidewall.
From top to bottom the histograms correspond to temperatures 10~K, 70~K, 90~K, and 160~K.}
\label{fig02}
\end{center}
\end{figure}

First, we collected the histograms depicting probabilities to find oxygens and hydrogens at a certain radius from the CNT axis (Fig.~\ref{fig02}).
The $x$'s fit the range of values from 0 (near axis) to 1 (near the CNT sidewall).
One can clearly see that neither oxygen, nor hydrogen atoms spotted at $x > 0.7$.
The radius values were calculated at regular periods (every 50 MD timesteps) for the atoms belonging to central water molecules,
except four terminal ones -- two from each side.
The oxygen histogram for the lowest temperature of 10~K (top panel) demonstrates a moderate peak centered at $x=0.2$.
The hydrogen histogram for $T=10$~K shows a sharper peak at $x=0.2$, then followed by a shoulder between $x=0.35$ and 0.55.
The oxygen histogram for $T=70$~K keeps the shape and slightly shifts to smaller $x$'s, while the hydrogen histogram shows one broad peak up to $x=0.6$.
The $T=90$~K, 160~K histograms look qualitatively similar to the $T=70$~K case:
They become broader, showing slight shifts of maxima for oxygens to smaller $x$'s, and for hydrogens -- to larger $x$'s.
In Fig.~\ref{fig02} we presented only the results for the CNT $\approx85$~{\AA} long, since the other cases resemble this behavior.
As no significant atom occupancy at $x=0$ is spotted,
it reflects the fact that none of the oxygens and hydrogens of water reside on the CNT axis.

\begin{figure}[htb!]
\begin{center}
\includegraphics[clip=true,width=0.9\columnwidth]{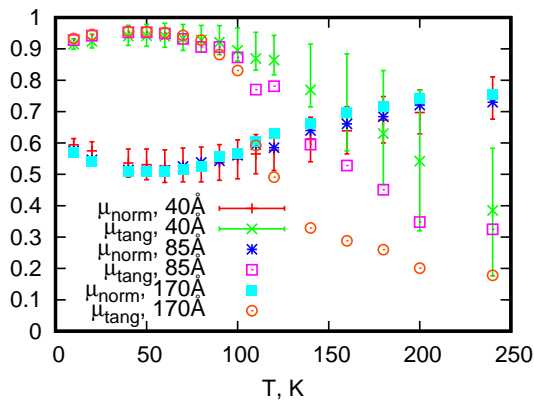}
\caption{The mean magnitude of the perpendicular dipole component for one water molecule $\mu_{\rm norm}$
(red, blue, cyan)
and
the mean total dipole moment along the CNT axis of central molecules in the chain divided by number of molecules $\mu_{\rm tang}$
(green, violet, orange).
Both quantities are additionally normalized by the magnitude of the dipole moment of an individual water molecule $\mu$.
Three CNT length cases considered: $\approx40$~{\AA}, $\approx85$~{\AA}, and $\approx170$~{\AA}.
We show error bars for the CNT of the length {$\approx 40$~\AA} only in order to avoid overcrowding the figure.}
\label{fig03}
\end{center}
\end{figure}

Next stage is aimed to reproduce results in Fig.~S8 from Supplemental Material for the paper of X.~Ma {\it et al.} \cite{Ma2017}.
Using the same definitions as X.~Ma {\it et al.}
we show in Fig.~\ref{fig03}
normal component of dipole moment of individual water molecules $\mu_{\rm norm} = \overline{\vert\mu_{\perp}\vert}/\mu$
and
total dipole moment of water chain tangential to the CNT axis $\mu_{\rm tang} = \overline{\mu^{\rm tot}_{\parallel}}/(N \mu)$;
here
$\overline{(\ldots)}$ denotes the mean value of $(\ldots)$.
For the sake of clarity, one needs to mention that these two quantities are not the components of the same vector:
The normal one is the mean over all normal components, wherever they point (a sum of modules of perpendicular components of dipole moments),
while the tangential one is the mean projection of the total dipole moment on the CNT axis (vector sum projected on the axis).
Fig.~\ref{fig03} presents three sets of datapoints for each of the CNT length case.
As the configuration snapshots were taken at regular time intervals,
we can calculate the instant values (normal/tangential components of dipole moment, water-water distances) for a given configuration at a given time step, then the averages over multiple time steps were done.
Having the instant values, we were also able to assess their variability along the simulation course.
In particular,
Fig.~3 and the lower panel of Fig.~4 show the simulation results
with the error bars included, indicating 25th and 75th percentiles.
Since Fig. 3 reports the results for a series of simulations for different CNT lengths,
the error bars are presented only for the case of 40~{\AA} long CNT in order to avoid an overcrowding on the plot
-- the remaining results demonstrate roughly the same level of variability.
A general trend is that the variability increases with the temperatures.
The graphs for $\mu_{\rm norm}$ show much less variation with the CNT length with a slightly larger amplitude for the longest CNT.
This is not the case for $\mu_{\rm tang}$:
All three graphs almost coincide for the temperatures below 100~K, followed by a strong deviation for higher temperatures.
The longer CNT the stronger decrease of $\mu_{\rm tang}$ is observed.
Both $\mu_{\rm tang}$ and $\mu_{\rm norm}$ reach maximum and minimum, correspondingly, at $T=50$~K.
The stronger decrease of the tangential components at $T>100$~K for the longer CNTs
we interpret as a gradual loss of high-temperature correlations with increase of the volume per water molecule.
X.~Ma and coauthors observed similar temperature profiles and pointed out three types of water arrangement:
1) hydrogen-bonding over the whole chain, when dipole moments of water molecules are tilted by 31$^\circ$ to the CNT axis,
2) dipole moments tend to align along the CNT axis in one direction,
and
3) collective arrangement is completely destroyed.
On the basis of Fig.~\ref{fig03}
we can assume
that the quasiphase 2 is achieved in vicinity of $T=50$~K, while the quasiphases 1 and 3 are located at lower and higher temperatures, respectively.
One has to note, that, despite the difference in water models (we used SPC/E, X.~Ma {\it et al.} -- TIP3P), the results show a semi-quantitative agreement.

The reported results are obtained within the model with optimized charges on oxygens and hydrogens of water.
Important to note that before the version with optimized charges
we also utilized the SPC/E model with original charges to tackle the problem.
These results are not reported here,
since this model is unable to reproduce the expected transitions between the above mentioned three quasiphases.
Turns out, the original SPC/E model reveals the hydrogen-bonding driven arrangement at low temperature,
which then evolves to a chaotic phase at higher temperatures.

\begin{figure}[htb!]
\begin{center}
\includegraphics[clip=true,width=0.9\columnwidth]{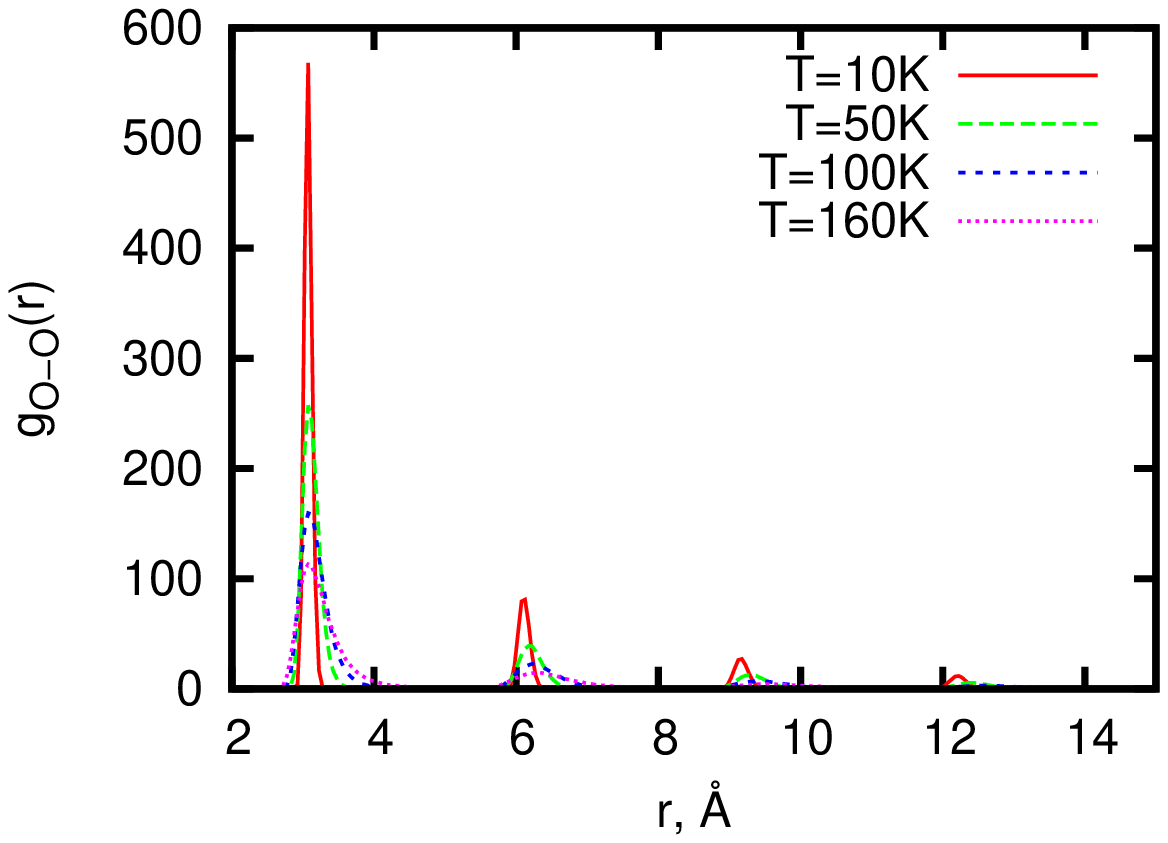}
\includegraphics[clip=true,width=0.9\columnwidth]{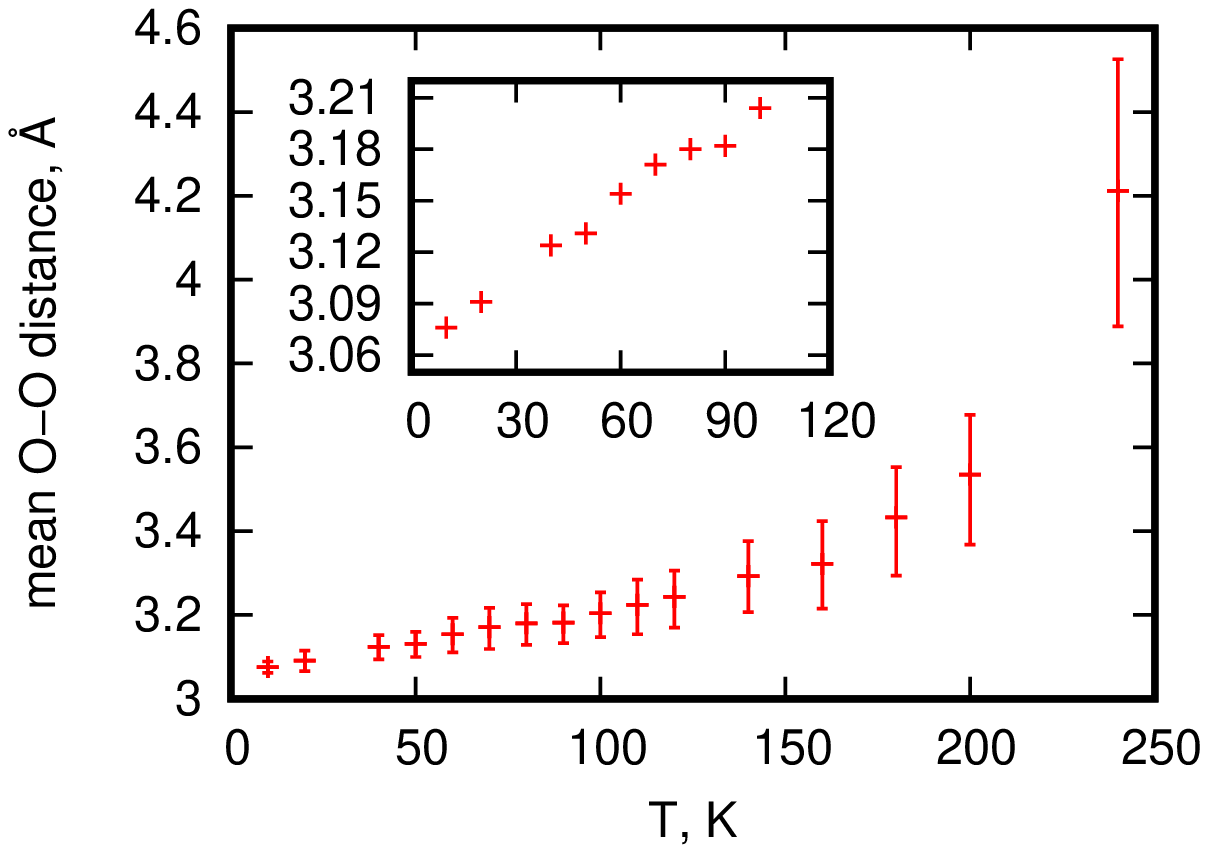}
\caption{(Top)
Radial distribution functions between oxygens of water $g_{\rm OO}(r)$
for four different temperatures $T=10, 50, 100, 160$~K
for CNT of the length $\approx40$~{\AA}.
(Bottom)
Mean distances between the nearest oxygen atoms within a water chain as a function of temperature
for CNT of the length $\approx40$~{\AA}.
The inset shows the enlarged view in the low-temperature region.}
\label{fig04}
\end{center}
\end{figure}

The MD results we presented so far allow us to conclude
that the setup utilized for simulations is able to catch the main features of dipolar rearrangement
as hinted by temperature-dependent photoluminescence spectroscopy experiments in Ref.~\cite{Ma2017}.
Therefore, the structural and energetic details drawn from the MD simulations can be used at further stages of a lattice model construction,
see Sec.~\ref{sec3}.
In particular, a mean distance between nearest water molecules is an important reference parameter.
For this purpose we calculated the radial distribution functions $g_{\rm OO}(r)$ (shown in the top panel of Fig.~\ref{fig04}).
The distributions reproduce the periodic nature of a water chain,
however, the positions of $g_{\rm OO}(r)$ peaks do not change, while the peaks become wider.
Therefore, we changed the approach to monitor the oxygen-oxygen distances during the simulation at regular periods and then averaged them
(shown in the bottom panel of Fig.~\ref{fig04}).
It is worth noting that although a rough estimate of the maximal oxygen-oxygen distance is $\approx40$~\AA$/11\approx3.64$~\AA,
the mean ${\rm O}{-}{\rm O}$ distance exceeds 4~{\AA} at high temperatures
evidencing that the water molecules lie on a zig-zag path rather than on a straight line,
see Fig.~\ref{fig01}.
The reported mean oxygen-oxygen distances may be also of use for understanding a behavior of water molecules in the CNT.

\section{Lattice model which accounts for interactions and rotations}
\label{sec3}

\subsection{Formulation of the model}
\label{sec31}

The quantum-chemical computations and MD simulations illustrated in Sec.~\ref{sec2}
allow us to suggest a simple model for behavior of water molecules forming a single-file chain in CNT.
The first ingredient of the model is the interactions.
Strong short-range nearest-neigh\-bor interactions resulting in bonding of water molecules into hydrogen-bonded chains
and
long-range dipole-dipole interactions
are important especially at low temperatures.
The second ingredient of the model is the rotations.
Since the diameter of CNT is around 0.5~nm so that mutual passage of water molecules is excluded,
important restrictions for rotations hold.
Namely,
water molecules can rotate as one entity only around the axis which is directed along the nanotube axis
if they form a hydrogen-bonded chain.
Besides,
a few linked molecules rotating as a whole have much less microstates than the same not connected molecules rotating independently.
Moreover,
each water molecule separated far enough from other water molecules can rotate around three axes as any rigid body in the three-dimensional space.
Rotations, which contribute to the entropy, are important especially at high temperatures.
An interplay of these two competing factors, interactions and rotations,
can produce a following temperature-driven collective behavior:
Interactions win at lower temperatures yielding hydrogen-bonded chains and a corresponding mean dipole moment along the nanotube axis,
which is not the maximum,
whereas
rotations win at higher temperatures yielding completely independent rotating water molecules with vanishing mean dipole moment;
most importantly,
at a wide range of intermediate temperatures,
the configurations with dipole moment along the nanotube axis statistically dominate
yielding a temperature-driven increase of the mean dipole moment along the nanotube.
We may implement this picture into a simple one-dimensional lattice model.
Lattice models are widely used for a discretized description of continuum fluids \cite{Stanley1971}.
A lattice chain of dipoles which interact and rotate is known in context of other studies, see, e.g., Ref.~\cite{Dolgikh2013}.

\begin{figure}
\begin{center}
\includegraphics[clip=true,width=0.9\columnwidth]{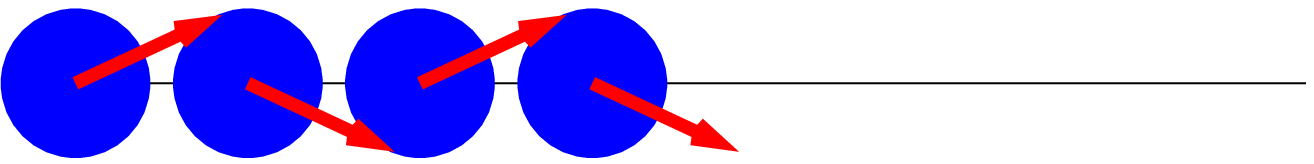}\\
\vspace{1mm}
\includegraphics[clip=true,width=0.9\columnwidth]{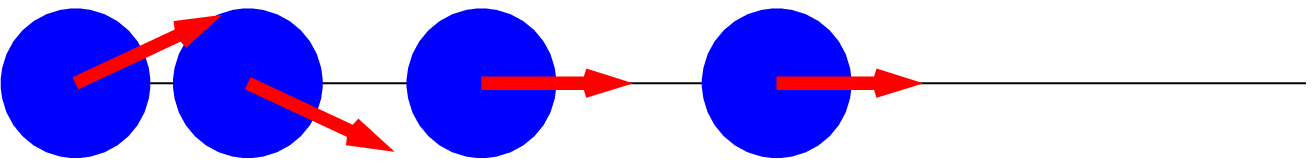}\\
\vspace{1mm}
\includegraphics[clip=true,width=0.9\columnwidth]{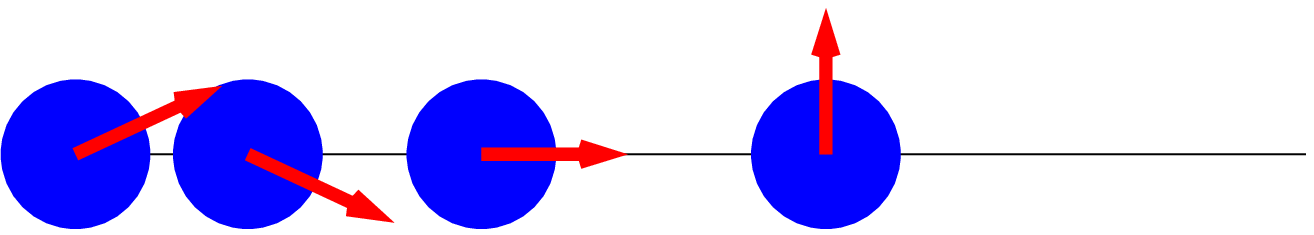}\\
\vspace{1mm}
\includegraphics[clip=true,width=0.9\columnwidth]{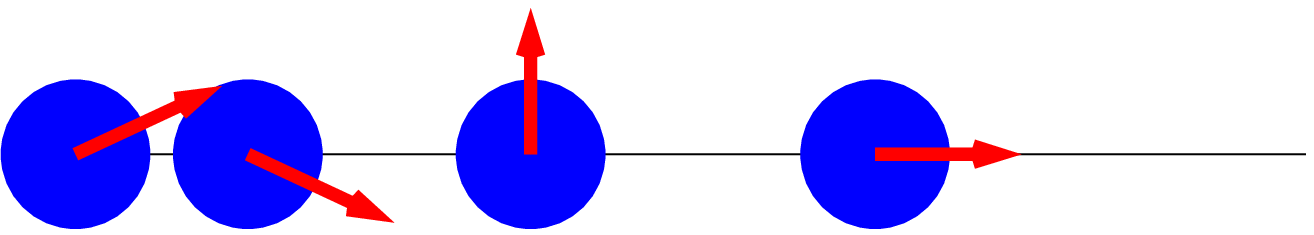}\\
\vspace{1mm}
\includegraphics[clip=true,width=0.9\columnwidth]{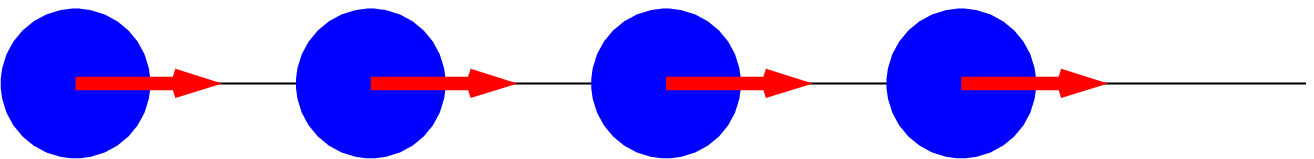}\\
\vspace{1mm}
\includegraphics[clip=true,width=0.9\columnwidth]{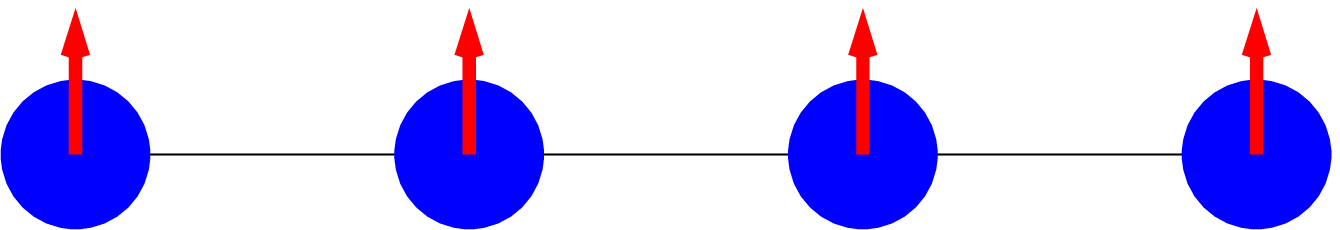}\\
\vspace{1mm}
\caption{From top to bottom:
Pictorial illustration for configurations
which correspond to the (allowed) states $1111$, $1122$, $1123$, $1132$, $2222$, and $3333$ of the $N=4$ lattice model.}
\label{fig05}
\end{center}
\end{figure}

\begin{figure}[htb!]
\begin{center}
\includegraphics[clip=true,width=0.9\columnwidth]{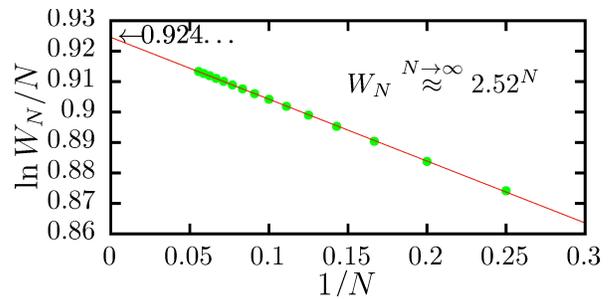}
\caption{Towards the number of states per site.
Total number of allowed states $W_N$ of the $N$-site model scales as $\approx 2.52^N$ when $N\to\infty$
since $\ln W_N/N\to0.924\,33\ldots$ when $N\to\infty$ (linear fit),
i.e., the model has $\approx 2.52$ states per each lattice site.}
\label{fig06}
\end{center}
\end{figure}

{\it Phase space and partition function}.
To grasp the described behavior of water in CNT,
we propose a simple lattice model with $3$ states at each lattice site subjected then to certain restrictions
which effectively decrease the number of states per site to $\approx 2.52$.
More specifically,
let us consider $N$ rigid bodies each with the moment of inertia $I$
(for the moments of inertia of a water molecule see Refs.~\cite{Bier2018,Sahoo2021}),
which for simplicity carry coplanar dipoles $\vec{\mu}_j$, $j=1,\ldots,N$;
they are rendered on a single straight line so that the distance between the neighboring sites $j$ and $j+1$ is $a_{j,j+1}$.
Moreover, we assume that each site $j$ may be in one of the following 3 states $\xi_j$:
\begin{itemize}
\item
The state $\xi_j=1$,
when $\mu_{\parallel,j}=\mu \cos \alpha_1$, $\vert\mu_{\perp,j}\vert=\mu \sin \alpha_1$,
and the extension of the occupied site is $a_1$ distributed symmetrically to the left ($a_1/2$) and to the right ($a_1/2$);
site being in such a state belongs to a hydrogen-bonded chain;
a set of $\mu_{\perp,j}$ for the hydrogen-bonded chain forms a staggered pattern;
we set $\alpha_1=31^{\circ}$ \cite{Ma2017};
\item
the state $\xi_j=2$,
when $\mu_{\parallel,j}=\mu \cos \alpha_2$, $\vert\mu_{\perp,j}\vert=\mu \sin \alpha_2$, $0^{\circ}\le\alpha_2<\alpha_1$,
and the extension of the occupied site is $a_2=a_1(1+\varepsilon_2)>a_1$;
we set $\alpha_2=0^{\circ}$;
\item
the state $\xi_j=3$,
$\mu_{\parallel,j}=\mu \cos \alpha_3$, $\vert\mu_{\perp,j}\vert=\mu \sin \alpha_3$, $\alpha_1<\alpha_3\le 90^{\circ}$,
and the extension of the occupied site is $a_3=a_1(1+\varepsilon_3)>a_2$;
within our essentially minimal description,
this state represents a completely independent water molecule with a random orientation of $\vec{\mu}_j$;
we set $\alpha_3=90^{\circ}$.
\end{itemize}
Furthermore,
\begin{itemize}
\item
the state $\xi_j=1$ indicating a hydrogen-bonded chain must have at least one neighboring site in the same state $\xi=1$,
otherwise such a configuration is forbidden.
That is, the states containing patterns like
$\ldots212\ldots,$ $\ldots213\ldots,$ $\ldots312\ldots,$ or $\ldots313\ldots$
are forbidden, see Fig.~\ref{fig05}.
\end{itemize}
The imposed restriction reduces the number of states $W_N$ for the lattice of $N$ sites, which is now smaller than $3^N$.
By inspection, we find
$W_4=33<81$,
$W_5=83<243$,
$W_6=209<729$,
$W_7=527<2\,187$,
$W_8=1\,329<6\,561$,
$W_9=3\,351<19\,683$,
$W_{10}=8\,449<59\,049$
etc.
Extrapolating to the thermodynamic limit $N\to\infty$ (linear fit), we obtain $W_{N}\approx 2.52^N$, see Fig.~\ref{fig06}.
That is, the introduced lattice model has $\approx 2.52<3$ states per each site.
We also remark that although we are interested in the finite-$N$ cases,
the imposed restriction preserves the correct thermodynamic behavior when $N\to\infty$, too, cf. Ref.~\cite{Stasyuk1992}.

We may note in passing that within our simple model
the independent hydrogen-bonded chains (finite clusters) are necessarily separated by at least one site in the state 2 or 3.

We are interested in thermodynamic quantities which are related to the partition function
\begin{eqnarray}
\label{01}
Z={\sum_{\xi_1\ldots\xi_N}}^{\prime}\sum_{\rm rot}\exp\left[-\frac{E(\xi_1\ldots\xi_N)}{k_{\rm B}T}\right].
\end{eqnarray}
Here
the prime near the first sum indicates the discussed above restriction on the set of values $\xi_1 \ldots \xi_N$
and
the second sum denotes the summation over rotational degrees of freedom for given (allowed) set $\xi_1 \ldots \xi_N$.
Moreover,
$E(\xi_1 \ldots \xi_N)$ stands for the sum of the rotation energy and the interaction energy
which contribute to
the rotation part $K$
and
the interaction part $Q$
of the partition function $Z=Z(T,N)$,
see, e.g., Eqs.~(\ref{10}) and (\ref{11}) below.

{\it Interactions}.
We take into account the short-range nearest-neighbor interactions
by treating the water mo\-lecules at sites $j$ and $j+1$ with $a_{j,j+1}=a_1$ as linked (i.e., rigidly connected) through a  hydrogen bond.
The long-range interactions between all water molecules $U_{\xi_1\ldots\xi_N}$
is the sum over all $N(N-1)/2$ pairs of the dipole-dipole interaction $u_{ij}$, $i<j$, $i=1,\ldots,N-1$, $j=2,\ldots,N$
(electrostatic interactions between charges in a metallic CNT might be more complicated,
see Refs.~\cite{Weber1939,Kornyshev2013,Rochester2013}).
Moreover,
\begin{eqnarray}
\label{02}
u_{ij}=k\frac{\mu_{\perp,i}\mu_{\perp,j}-2\mu_{\parallel,i}\mu_{\parallel,j}}{a^3_{ij}},
\;\;\;
k=\frac{1}{4\pi\epsilon_0}
\end{eqnarray}
[$\epsilon_0$ is the vacuum permittivity (SI units)],
if the both sites $i$ and $j$ belong to the same hydrogen-bonded chain.
However,
\begin{eqnarray}
\label{03}
u_{ij}=k\frac{-2\mu_{\parallel,i}\mu_{\parallel,j}}{a^3_{ij}},
\end{eqnarray}
if the sites $i$ and $j$ belong to different hydrogen-bonded chains or at least one of these sites is in the state 2 or 3.
In other words,
the $\mu_{\perp}$ on-site components contribute to the intersite interaction $u_{ij}$ only if the both sites rotate as a whole,
but do not contribute to the intersite interaction if they rotate independently.
In contrast,
the $\mu_{\parallel}$ on-site components always contribute to the intersite interaction $u_{ij}$.

{\it Rotations}.
Our simple description of limited rotations is as follows.
A hydrogen-bonded chain consisting of $n$ water molecules has the moment of inertia $nI$ and rotates along one axis only,
which coincides with the nanotube axis.
Its energy is given by $E_m=\hbar^2 m^2/(2nI)$ with $m=0,\pm 1,\pm 2,\ldots$
and hence each energy level $E_m$ except the one with $m=0$ is two-fold degenerate \cite{Galitski1981}.
The rotational partition function of the hydrogen-bonded chain reads:
\begin{eqnarray}
\label{04}
K_{n}^{(1)}=\sum_{m=-\infty}^\infty\exp\left(-\frac{m^2}{n\tau}\right)
\overset{\tau>1}{\longrightarrow}\sqrt{\pi\tau},
\nonumber\\
\tau=\frac{T}{T_{\rm rot}},
\;
k_{\rm B}T_{\rm rot}=\frac{\hbar^2}{2I}.
\end{eqnarray}
Furthermore,
we assume that a site being in the state 2 corresponds to a water molecule which rotates similarly to the hydrogen-bonded chain,
i.e., contributes $K_{1}^{(1)}$ to the rotation part of the partition function.
Moreover,
a site being in the state 3 corresponds to a water molecule which rotates along three axes;
its energy is given by $E_J=\hbar^2 J(J+1)/(2I)$ with $J=0,1,2,\ldots$
and the degeneracy of the energy level $E_J$ is $(2J+1)^2$.
The partition function of such a rotor reads:
\begin{eqnarray}
\label{05}
K_{1}^{(3)}=\sum_{J=0}^\infty\left(\left(2J+1\right)^2\exp\left[-\frac{J(J+1)}{\tau}\right]\right)
\nonumber\\
\overset{\tau>1}{\longrightarrow}\sqrt{\pi\tau^3}\exp\frac{1}{4\tau}.
\end{eqnarray}

Interestingly,
$K_{n}^{(1)}$ (\ref{04}) and $K_{1}^{(3)}$ (\ref{05}) are related to the theta functions
$\theta_2(v,\varkappa)=\sum_{n=-\infty}^{\infty}q^{(n-1/2)^2}{\rm e}^{(2n-1)\pi v{\rm i}}$
and
$\theta_3(v,\varkappa)=\sum_{n=-\infty}^{\infty}q^{n^2}{\rm e}^{2n\pi v{\rm i}}$
with $q={\rm e}^{-\pi\varkappa}$ \cite{Jahnke1960}.
Namely,
$K_{n}^{(1)}=\theta_3(0,\varkappa)$, $\varkappa=1/(\pi n\tau)$
and
$K_{1}^{(3)}=2\tau^2{\rm e}^{1/(4\tau)}{\rm d}\theta_2(0,\varkappa)/{\rm d}\tau$, $\varkappa=1/(\pi\tau)$.

{\it Quantities of interest}.
Having the partition function $Z$ given in Eq.~(\ref{01})
we immediately get the Helmholtz free energy $F=-k_{\rm B}T\ln Z$ and hence
the entropy $S=-\partial F/\partial T$,
the internal energy $E=F+TS$,
and the specific heat $C=T\partial S/\partial T$.

Moreover,
the thermodynamic average is defined as follows:
\begin{eqnarray}
\label{06}
\langle(\ldots)\rangle
=\frac{1}{Z}
{\sum_{\xi_1\ldots\xi_N}}^{\prime}\sum_{\rm rot}
\left(\exp\left[-\frac{E(\xi_1 \ldots \xi_N)}{k_{\rm B}T}\right](\ldots)\right).
\end{eqnarray}
We are interested in the average dipole moment (per site)
or more precisely in the following quantities:
\begin{eqnarray}
\label{07}
\mu_{\parallel}=\frac{1}{N}\sum_{j=1}^N\langle\mu_{\parallel,j}\rangle,
\;\;\;
\vert\mu_{\perp}\vert=\frac{1}{N}\sum_{j=1}^N\langle\vert\mu_{\perp,j}\vert\rangle.
\end{eqnarray}
Obviously,
$\mu_{\rm tang}=\mu_{\parallel}/\mu$
and
$\mu_{\rm norm}=\vert\mu_{\perp}\vert/\mu$,
see Sec.~\ref{sec22} and Fig.~\ref{fig03}.
The dipole correlations are defined as follows:
\begin{eqnarray}
\label{08}
\langle\vec{\mu}_i\cdot\vec{\mu}_{j}\rangle
=
\langle\mu_{\parallel,i}\mu_{\parallel,j}\rangle+\langle\mu_{\perp,i}\mu_{\perp,j}\rangle
\end{eqnarray}
and $\sqrt{\langle\vec{\mu}_j\cdot\vec{\mu}_{j}\rangle}$ is the average dipole moment at the site $j$.

Finally,
we can calculate
the average length of the chain $L$
and
the coefficient of linear thermal expansion $\alpha_L=(1/L)(\partial L/\partial T)$:
\begin{eqnarray}
\label{09}
L=\sum_{j=1}^{N-1}\langle a_{j,j+1}\rangle,
\;\;\;
\alpha_L=\frac{1}{L}\frac{{\rm d} L}{{\rm d} T}.
\end{eqnarray}
The length of the chain per site $L/N$ might be related to the mean distance between the nearest oxygen atoms,
see the lower panel of Fig.~\ref{fig04}.

{\it Scales and units}.
There are only a few quantities which are used as the input for the lattice model described above.
We begin with the length scale determined by $a_1$ which is about $3$~{\AA}.
We assume for simplicity that $a_2=(1+\varepsilon)a_1$ and $a_3=(1+2\varepsilon)a_1$ with $\varepsilon=0.08$.
Importantly, the value of $\varepsilon$ must exceed a certain threshold value
in order to have as the ground state the hydrogen-bonded chain $111\ldots$ rather than the state $22\ldots2$.
Next, the energy scales are determined by $T_{\rm rot}$ given in Eq.~(\ref{04}) and $T_{\rm dip}=(k\mu^2/a^3)/k_{\rm B}$.
Using for $I$ the values $1.0,2.9,1.9\times 10^{-47}$ in units of SI \cite{Bier2018}
we obtain for $T_{\rm rot}$ the values $39,14,21$~K.
In our calculations we set $T_{\rm rot}=20$~K.
Another energy scale $T_{\rm dip}$ depends on the values of the dipole moment $\mu$ and the characteristic length $a$.
For simplicity, we set $T_{\rm dip}=200$~K, that is  $T_{\rm dip}=T_{\rm rot}/R$ with $R=0.1$.
Last but not least, we remind that $\alpha_1=31^\circ$, $\alpha_2=0^\circ$, $\alpha_3=90^\circ$ have been assumed above.
This choice agrees with MD simulations.

It is important to stress that all the results presented below depend only quantitatively on the chosen parameters
$a_j$, $\alpha_j$, $j=1,2,3$, $T_{\rm rot}$, and $R=T_{\rm rot}/T_{\rm dip}$,
i.e., all conclusions are robust and do not require a fine tuning of the input parameters.
Moreover,
they are in a reasonable agreement with the ones used in MD simulations.
Thus, the MD results imply that $a_2/a_1$ exceeds 1 by about $0.026\ldots0.033$,
see the bottom panel of Fig.~\ref{fig04}.
Assuming $\mu=1.105$~D and $a$ in the range $3.05\ldots3.10$~{\AA} we arrive at $T_{\rm dip}$ about $300$~K
which, for $T_{\rm rot}=20$~K, corresponds to $R\approx0.07$.
We emphasize here that our aim is not to reproduce MD simulations,
which present a rough classical cartoon for clarifying experimental observations,
but only to illustrate the ability of the introduced lattice model to mimic quasiphases in a single file of water molecules in CNT
as they were discussed in Ref.~\cite{Ma2017}.
After all,
$a_j$, $\alpha_j$, $j=1,2,3$, $T_{\rm rot}$, and $R=T_{\rm rot}/T_{\rm dip}$ can be also viewed as fitting parameters.

\subsection{Properties of the model}
\label{sec32}

As discussed above, we use for concreteness the set of parameters
$\alpha_1=31^\circ$, $\alpha_2=0^\circ$, $\alpha_3=90^\circ$,
$a_1=3$~\AA, $a_2=1.08a_1$, $a_3=1.16a_1$,
$T_{\rm rot}=20$~K, and $R=0.1$.
We perform all calculation using the Maple software package implemented on a personal computer.

To illustrate how an interplay between interactions and rotations can produce an intriguing temperature-driven cooperative behavior,
we begin with consideration of the case of $N=4$ site lattice model,
when the phase space contains 33 states.
We have checked by inspection that among these 33 states contributing to the partition function $Z$ in Eq.~(\ref{01})
only 1 term  ($1111$) contains $K_4^{(1)}$
corresponding to the hydrogen-bonded chain of length 4,
4 terms ($1112$, $1113$, $2111$, $3111$) contain $K_3^{(1)}$
(hydrogen-bonded chains of length 3),
12 terms ($1122$, \ldots, $3311$) contain $K_2^{(1)}$
(hydrogen-bonded chains of length 2),
and
the rest 16 terms ($2222$, \ldots, $3333$) contain contribution due to rotation of separate water molecules.
That is, the partition function (\ref{01}) reads:
\begin{eqnarray}
\label{10}
Z=K_4^{(1)} Q_{1111}+K_3^{(1)} {\cal Q}_3+K_2^{(1)} {\cal Q}_2+{\cal Z}_0,
\end{eqnarray}
where
$Q_{\xi_1\xi_2\xi_3\xi_4}=\exp[-U_{\xi_1\xi_2\xi_3\xi_4}/(k_{\rm B}T)]$
is the interaction contribution to the Gibbs factor from the state $\xi_1,\xi_2,\xi_3,\xi_4$
[see Eq.~(\ref{01})]
and
\begin{widetext}
\begin{eqnarray}
\label{11}
{\cal Q}_3=K_1^{(1)}\left(Q_{1112}+Q_{2111}\right)+K_1^{(3)}\left(Q_{1113}+Q_{3111}\right),
\nonumber\\
{\cal Q}_2=
\left(K_1^{(1)}\right)^2
\left(Q_{1122}+Q_{2112}+Q_{2211}\right)
\nonumber\\
+K_1^{(1)}K_1^{(3)}
\left(
Q_{1123}+Q_{1132}+Q_{2113}
+Q_{2311}+Q_{3112}+Q_{3211}
\right)
\nonumber\\
+\left(K_1^{(3)}\right)^2
\left(Q_{1133}+Q_{3113}+Q_{3311}\right),
\nonumber\\
{\cal Z}_0=
\left(K_1^{(1)}\right)^4Q_{2222}
+\left(K_1^{(1)}\right)^3K_1^{(3)}
\left(Q_{2223}+Q_{2232}+Q_{2322}+Q_{3222}\right)
\nonumber\\
+\left(K_1^{(1)}\right)^2\left(K_1^{(3)}\right)^2
\left(Q_{2233}+Q_{2323}+Q_{2332}+Q_{3223}+Q_{3232}+Q_{3322}\right)
\nonumber\\
+K_1^{(1)}\left(K_1^{(3)}\right)^3
\left(Q_{2333}+Q_{3233}+Q_{3323}+Q_{3332}\right)
+\left(K_1^{(3)}\right)^4Q_{3333}.
\end{eqnarray}
\end{widetext}
Here,
$K_3^{(1)}{\cal Q}_3$ is the contribution to the partition function $Z$ (10)
from all configurations with hydrogen-bonded chains of length $3$,
$K_2^{(1)}{\cal Q}_2$ is the contribution to the partition function $Z$ (10)
from all configurations with hydrogen-bonded chains of length $2$,
whereas ${\cal Z}_0$ is the contribution to the partition function $Z$ (10)
from all configurations without hydrogen-bonded chains.

\begin{figure}[htb!]
\begin{center}
\includegraphics[clip=true,width=0.8\columnwidth]{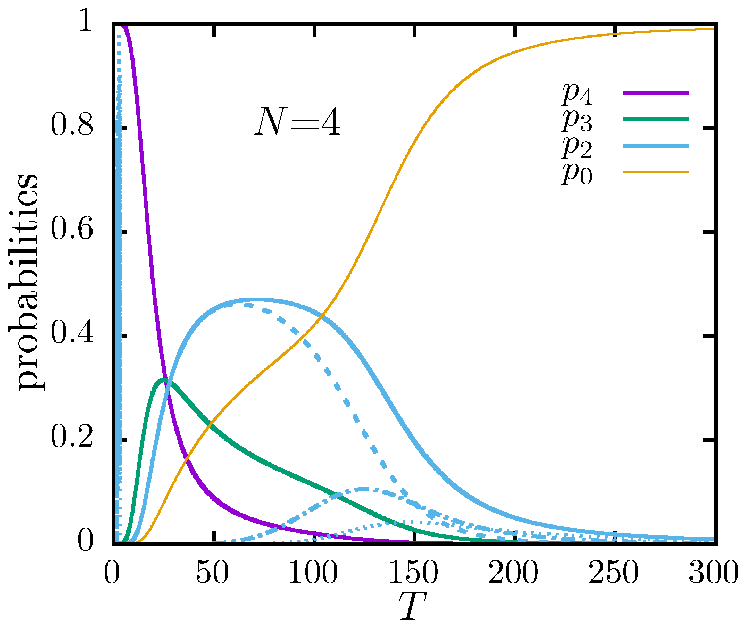}\\
\includegraphics[clip=true,width=0.8\columnwidth]{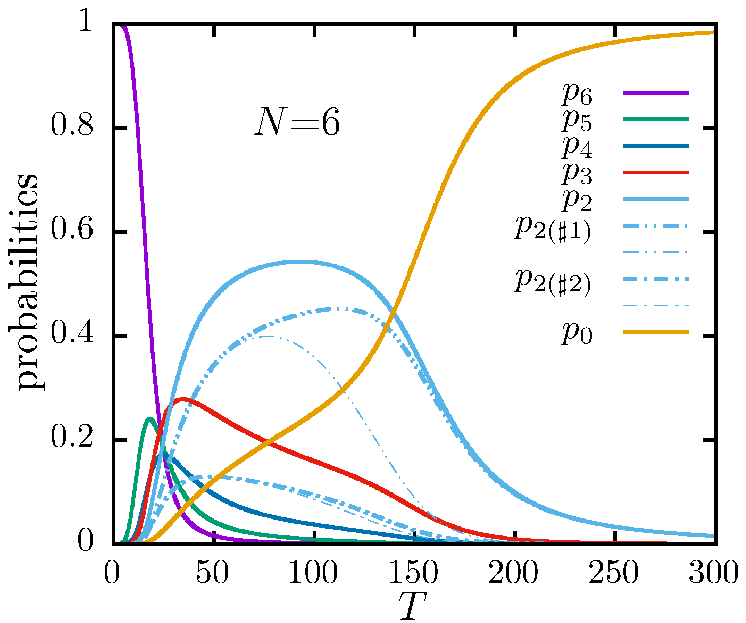}\\
\includegraphics[clip=true,width=0.8\columnwidth]{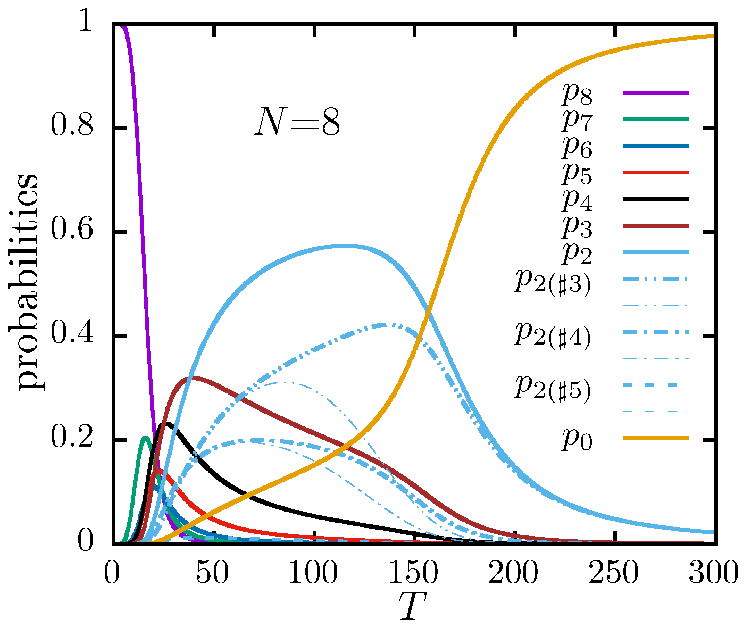}
\caption{Probabilities of various configurations versus temperature for the chains of (from top to bottom) $N=4,6,8$ sites.
Different broken blue curves illustrate different contributions to $p_2$;
for detailed explanation see main text.}
\label{fig07}
\end{center}
\end{figure}

It is worthwhile to introduce the probabilities
\begin{eqnarray}
\label{12}
p_4\!=\!\frac{K_4^{(1)}\! Q_{1111}}{Z},
p_3\!=\!\frac{K_3^{(1)}\! {\cal Q}_3}{Z},
p_2\!=\!\frac{K_2^{(1)}\! {\cal Q}_2}{Z},
p_0\!=\!\frac{{\cal Z}_0}{Z},
\nonumber\\
p_4+p_3+p_2+p_0=1.
\end{eqnarray}
The temperature-dependent probabilities $p_4$, $p_3$, $p_2$, $p_0$
control the role of the configurations with different length of hydrogen-bonded chains in thermodynamics.
In the zero-temperature limit $T\to 0$,
when the lowest-energy ground state dominates,
$Z\to K_4^{(1)}Q_{1111}$,
and $p_4\to1$.
In the high-temperature limit $T\to\infty$,
when the dipole-dipole interactions become irrelevant and $Q_{\xi_1\xi_2\xi_3\xi_4}\to 1$,
$Z\to K_4^{(1)}+2K_3^{(1)}(K_1^{(1)}+K_1^{(3)})+3K_2^{(1)}(K_1^{(1)}+K_1^{(3)})^2+(K_1^{(1)}+K_1^{(3)})^4
\to (K_1^{(3)})^4$,
and $p_0\to 1$.
Temperature dependencies of $p_4$, $p_3$, $p_2$, and $p_0$ (\ref{12}) are shown in the top panel of Fig.~\ref{fig07}.
As can be seen from this figure,
there is a wide temperature range of $40\ldots100$~K where the largest probability $p_2$ exceeds 40\%.
More detailed analysis of ${\cal Q}_2$ given in Eq.~(\ref{11}) shows
that the main contribution to $p_2$ below 150~K comes from the subset of configurations in which the two remaining molecules are in the state 2
(dashed blue line in the top panel of Fig.~\ref{fig07}),
but above 185~K the subset of configurations in which the two remaining molecules are in the state 3 becomes dominant
(dotted blue line in the top panel of Fig.~\ref{fig07}).
The subset of configurations in which the two remaining molecules are in the different states 2 and 3
although dominates for the temperature range 150\ldots185~K
(dash-dotted blue line in the top panel of Fig.~\ref{fig07}),
are still comparable with the two other contributions.

Let us pass to the case of the $N=6$ site lattice model (middle panel of Fig.~\ref{fig07}).
We again introduce the probabilities
$p_6$, $p_5$, $p_4$, $p_3$, $p_2$, and $p_0$
which contain
$K_6^{(1)}$, $K_5^{(1)}$, $K_4^{(1)}$, $K_3^{(1)}$, $K_2^{(1)}$, and $K_1^{(1)}$ or $K_1^{(3)}$,
respectively,
cf. Eqs.~(\ref{12}) and (\ref{11});
$p_6+p_5+p_4+p_3+p_2+p_0=1$.
Here,
$p_{3}$ contains
the terms with $K_3^{(1)}K_2^{(1)}$
(which describe configurations with two independent hydrogen-bonded chains of length 3 and 2 and one separate water molecule)
as well as
the terms with $K_3^{(1)}$
(which describe configurations with hydrogen-bonded chains of length 3 and the rest three separate water molecules).
Similarly,
$p_{2}$ contains
\begin{itemize}
\item
the terms with $(K_2^{(1)})^2$
(two independent hydrogen-bonded chains of length 2 and two separate water molecules;
their contribution $p_{2(\#2)}\equiv p_{2(+2+1+1)}$ to $p_2$ is shown by dash-doted blue line in the middle panel of Fig.~\ref{fig07})
\end{itemize}
as well as
\begin{itemize}
\item
the terms with $K_2^{(1)}$
(hydrogen-bonded chains of length 2 and four separate water molecules;
their contribution $p_{2(\#1)}\equiv p_{2(+1+1+1+1)}$ to $p_2$ is shown by dash-dot-doted blue line in the middle panel of Fig.~\ref{fig07}).
\end{itemize}
Temperature dependence of probabilities $p_6, \ldots, p_0$,
which illustrates the role of configurations with different numbers and lengths of hydrogen-bonded chains in thermodynamics,
is shown in the middle panel in Fig.~\ref{fig07}.
Again, within the temperature range of $60\ldots125$~K the largest probability is $p_2$ exceeding 50\%, see the blue solid line.
The main contribution to $p_2$ comes from the subset of configurations in which the four remaining molecules are in the state 2
(thin dash-dot-doted blue line in the middle panel of Fig.~\ref{fig07});
the subset of configurations with two independent hydrogen-bonded chains of length 2 and two remaining molecules in the state 2 is noticeably smaller
(thin dash-doted blue line in the middle panel of Fig.~\ref{fig07}).

The results for the case of $N=8$ site lattice model reported in the bottom panel in Fig.~\ref{fig07},
demonstrate the same properties of the probabilities $p_8$, \ldots, $p_2$, and $p_0$.
Namely, within roughly the same temperature range, $65\ldots145$~K,
the largest probability is $p_2$ exceeding 50\%, see the blue solid line.
Furthermore, $p_{2}$ contains
\begin{itemize}
\item
the terms with $(K_2^{(1)})^3$
(three independent hydrogen-bonded chains of length 2 and two separate water molecules;
their very small contribution $p_{2(\#5)}\equiv p_{2(+2+2+1+1)}$ to $p_2$ is shown by dashed blue line in the bottom panel of Fig.~\ref{fig07}),
\item
the terms with $(K_2^{(1)})^2$
(two hydrogen-bonded chains of length 2 and four separate water molecules;
their contribution $p_{2(\#4)}\equiv p_{2(+2+1+1+1+1)}$ to $p_2$ is shown by dash-doted blue line in the bottom panel of Fig.~\ref{fig07}),
\end{itemize}
as well as
\begin{itemize}
\item
the terms with $K_2^{(1)}$
(hydrogen-bonded chains of length 2 and six separate water molecules;
their contribution $p_{2(\#3)}\equiv p_{2(+1+1+1+1+1+1)}$ to $p_2$ is shown by dash-dot-doted blue line in the bottom panel of Fig.~\ref{fig07}).
\end{itemize}
From the bottom panel of Fig.~\ref{fig07} one immediately concludes
that the main contribution to $p_2$ comes from
the subset of configurations in which the six remaining molecules are in the state 2
(thin dash-dot-doted blue line)
and
the subset of configurations with two independent hydrogen-bonded chains of length 2 and the four remaining molecules in the state 2
(thin dash-doted blue line).

In summary,
the considered cases $N=4,\,6,\,8$
(probabilities $p_N,\ldots,p_0$ for $N=10,\,12,\,14,\,16$ are reported in Supplementary Material)
provide evidence that the states,
which contain water molecules in the on-site states 1 and 2,
are the most relevant ones in the temperature range $80\ldots160$~K
and result in the emergence of an intermediate quasiphase.
Since the water molecules in the state 2 and in the short hydrogen-bonded chains contribute to $\mu_{\parallel}$ but not to $\mu_{\perp}$,
the tangential/normal component of total dipole moment should increase/decrease in this temperature interval.
We have further evidence for that in the temperature dependencies of observable quantities to be discussed below.

The intermediate quasiphase is stable in a rather wide temperature region.
A rough estimate for temperatures of quasiphase transitions $T_1<T_2$ follows by equating the corresponding probabilities,
that is,
\begin{eqnarray}
\label{13}
p_3(T_1)=p_2(T_1),
\;\;\;
p_2(T_2)=p_0(T_2).
\end{eqnarray}
For the chosen set of parameters we get $T_1=28,\,28,\,38$~K and $T_2=104,\,139,\,157$~K as $N=4,\,6,\,8$.
(For longer chains with $N=10,\,12,\,14,\,16$
we have
$T_1=46,\,56,\,67,\,78$~K
and
$T_2=167,\,175,\,180,\,185$~K,
respectively.)

Interestingly, the very existence of the intermediate quasiphase is robust to small deviations of the chosen set of parameters.
While the short-range interactions leading to forming hydrogen-bonded chains are mounted into the model,
the relative contributions of rotations and long-range dipole interactions are controlled by the value of $R=T_{\rm rot}/T_{\rm dip}$:
Rotations dominate for $R\to\infty$ and long-range dipole interactions dominate for $R\to 0$.
For large $R$, the intermediate quasiphase shows up within quite a narrow region at low temperature
($T_{\rm rot}$ is fixed).
As $R$ decreases, the dipole interactions stabilize this quasiphase,
i.e., extend the region of its existence and push it to higher temperatures.

\begin{figure}[htb!]
\begin{center}
\includegraphics[clip=true,width=0.9\columnwidth]{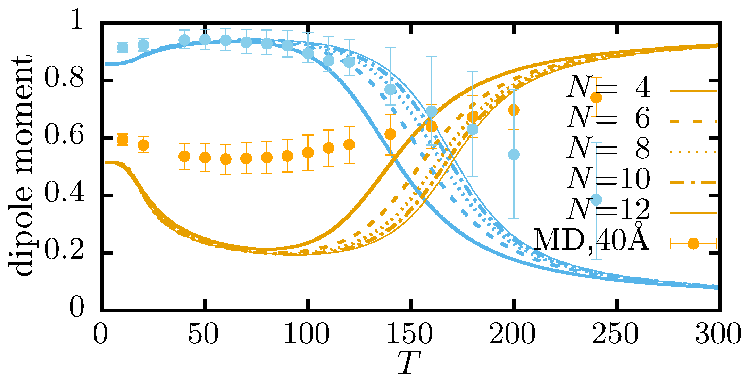}\\
\includegraphics[clip=true,width=0.9\columnwidth]{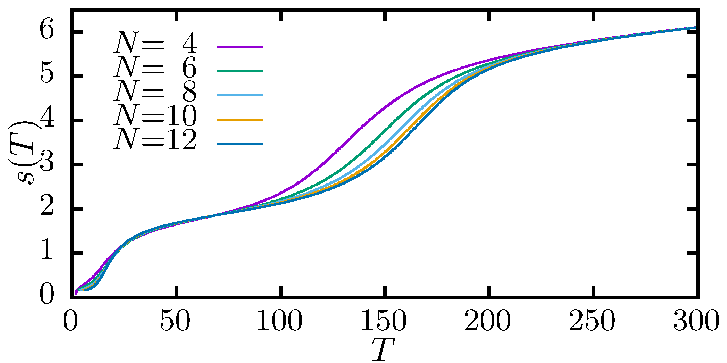}\\
\includegraphics[clip=true,width=0.9\columnwidth]{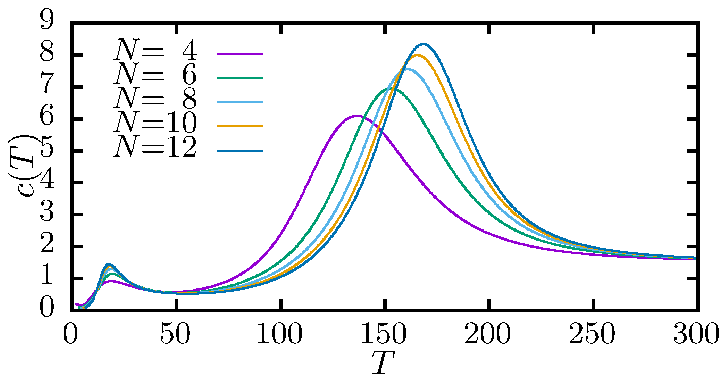}\\
\includegraphics[clip=true,width=0.9\columnwidth]{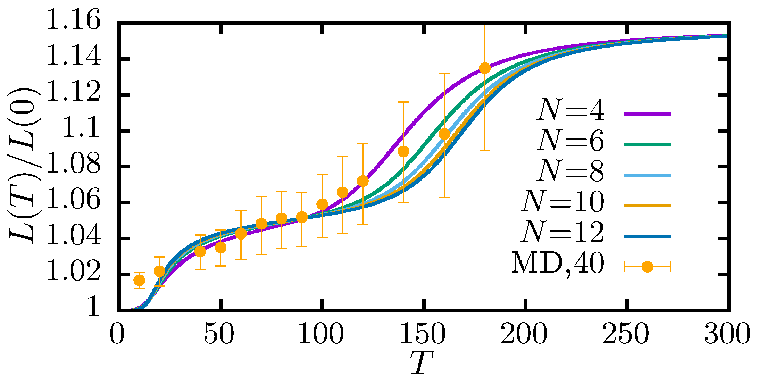}\\
\includegraphics[clip=true,width=0.9\columnwidth]{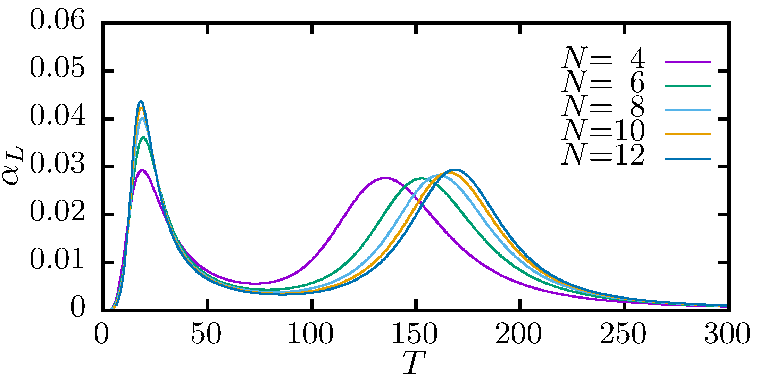}
\caption{Temperature dependencies of
(from top to bottom)
$\mu_{\parallel}$ (blue) and $\vert\mu_{\perp}\vert$ (orange),
entropy,
specific heat,
average length,
and
coefficient of linear thermal expansion
for the chains of $N{=}4,6,8,10,12$ sites.
MD simulations (CNT of length 40~{\AA}, filled circles) are shown in the panels
with $\mu_{\parallel}$, $\vert\mu_{\perp}\vert$ and the average length.}
\label{fig08}
\end{center}
\end{figure}

\begin{figure}[htb!]
\begin{center}
\includegraphics[clip=true,width=0.9\columnwidth]{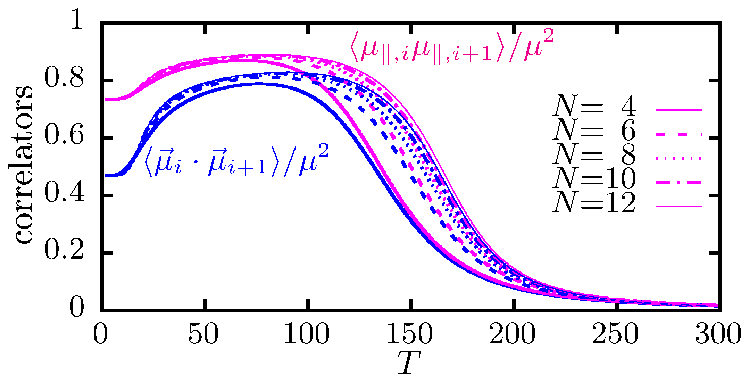}\\
\includegraphics[clip=true,width=0.9\columnwidth]{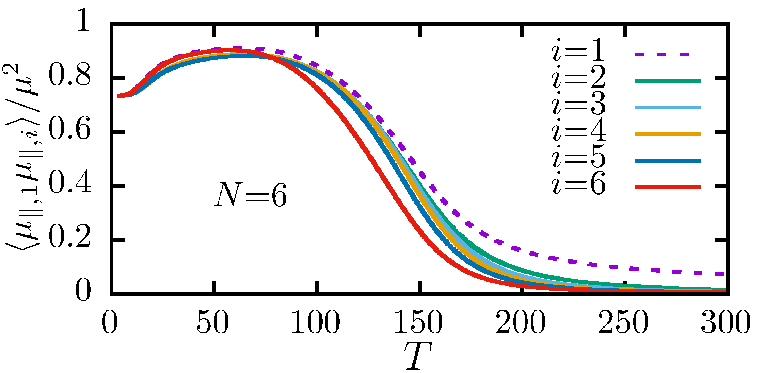}\\
\includegraphics[clip=true,width=0.9\columnwidth]{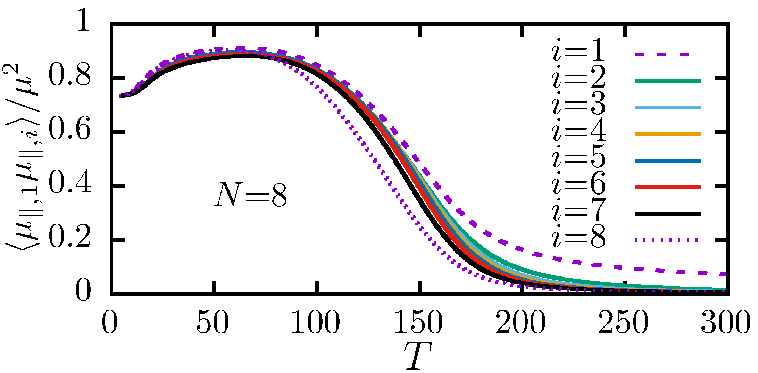}
\caption{(Top)
Nearest-neighbor correlators
$\sum_{i=1}^{N-1}\langle \vec{\mu}_i\cdot\vec{\mu}_{i+1}\rangle/[\mu^2(N-1)]$
and
$\sum_{i=1}^{N-1}\langle \mu_{i,\parallel}\mu_{i+1,\parallel}\rangle/[\mu^2(N-1)]$
(see MD simulation data in Fig.~S7 from Supplemental Material of Ref.~\cite{Ma2017}).
(Middle and bottom)
Correlators $\langle \mu_{\parallel,1} \mu_{\parallel,j} \rangle/\mu^2$, $j=1,\ldots,N$
for the chain of $N=6$ and $N=8$ sites.}
\label{fig09}
\end{center}
\end{figure}

In Figs.~\ref{fig08} and \ref{fig09} we show temperature dependencies for various quantities of interest for the lattice model of $N=4,\dots,12$ sites.
Note that even though the results converge as $N$ increases,
the difference between the cases $N=10$ and $N=12$ is still clearly seen.
First of all, we emphasize that the introduced model predicts
an increase of $\mu_{\parallel}$ (\ref{07}) and decrease of $\vert \mu_{\perp}\vert$ (\ref{07}) in the temperature range $20\ldots 100$~K
in agreement with MD simulations and interpretation of experimental data \cite{Ma2017},
see blue and orange curves in the top panel of Fig.~\ref{fig08}.
Although there are some finite-size effects,
the existence of increase of $\mu_{\parallel}$ and decrease of $\vert \mu_{\perp}\vert$ cannot be questioned.
We plot also the results of MD simulations for the CNT of length $\approx40$~{\AA} to illustrate qualitative agreement between both results.
We have to remark here that in MD simulations we face water molecules lying along a zig-zag path rather than along a straight line (see Fig.~\ref{fig01})
and this circumstance has also an impact on a visible difference between symbols and lines in the top panel of Fig.~\ref{fig08}.

A $\ln\!\sqrt{T}$-like part (up to a constant) in the temperature profile of entropy within $20\ldots100$~K
is replaced by a $\ln\!\sqrt{T^3}$-like part (up to a constant) in the temperature profile of entropy above $200$~K,
see the corresponding panel of Fig.~\ref{fig08}.
As it follows from Eqs.~(\ref{04}) and (\ref{05}),
these dependencies are the high-temperature behavior of rotators with one axis or three axes, respectively,
and hence $s(T)$ goes hand in hand
with leading contributions $p_{2(+1+\ldots)}$ to the dominant probability $p_2$ for intermediate temperatures
and
with the dominant probability $p_0$ for high temperatures.

The specific heat per site in the interval $20\ldots100$~K has values close to $k_{\rm B}/2$
signaling about separately rotating water molecules around the nanotube axis,
see Eq.~(\ref{04}).
At high temperatures it approaches $3k_{\rm B}/2$ as it should for independent three-axes rotators,
see Eq.~(\ref{05}).

The average length $L$ increases with the temperature growth, however, differently at different temperatures.
Two lower panels of Fig.~\ref{fig08}
illustrate a weak temperature dependence of the average chain length $L$ and small values of $\alpha_L$ (\ref{09})
within the temperature range $50\ldots100$~K.
We plot also the results of MD simulations shown previously in the lower panel of Fig.~\ref{fig04} after assuming $L(0)/(N-1)=3.025$~{\AA}.
Again both results, lines and filled circles, agree qualitatively
varying quite similarly between their minimal ($T\to0$) and maximal ($T\to\infty$) values.

In the upper panel of Fig.~\ref{fig09} we report the lattice-model predictions for the nearest-neighbor dipole correlators
[see Eq.~(\ref{08})],
which are presented in Fig.~S7 from Supplemental Material of Ref.~\cite{Ma2017}.
We notice here that
$\langle\mu_{\perp,i}\mu_{\perp,i+1}\rangle=0$ if the sites $i$ and $i+1$ do not belong to the same hydrogen-bonded chain
and
$\langle\mu_{\perp,i}\mu_{\perp,i+1}\rangle=-\sin^2 31^{\circ}\approx-0.265$ otherwise.

Moreover, from Fig.~\ref{fig09},
the correlations are almost independent on distance between the sites up to about $120$~K
(the lines corresponding to $i=2,\ldots,N-1$ are almost indistinguishable),
that indicates a correlated state of water molecules in CNT.
For higher temperatures,
the dipole correlations decrease with increase of the distance between sites.

Finally, rough estimates of the temperature interval for the intermediate quasiphase,
as they follow from inspection of various quantities,
are slightly different and depend on the quantity under analysis.
This is yet another indication that the gradual emergence of the intermediate quasiphase is not a strict phase transition.

\section{Conclusions}
\label{sec4}

In conclusion,
motivated by the suggestion of the experimental paper by X.~Ma {\it et al.} \cite{Ma2017},
we have performed quantum chemistry calculations as well as MD simulations for water molecules encapsulated in (6,5) CNT
to demonstrate how a temperature-driven dipole ordering shows up.
MD simulation outcomes depend on the input characteristics of the water molecule inside CNT.
The tangential (normal) component of total dipole moment
[i.e., $\mu_{\rm tang}$ ($\mu_{\rm norm}$)]
has a maximum (minimum) at the intermediate temperatures
if the values of water oxygen and hydrogen charges are significantly smaller than those commonly used within the TIP3P or SPC/E water models.
On the other hand, the outcomes of quantum chemistry calculations depend on a specific method employed
and, therefore, are not fully conclusive remaining an important issue to be resolved in the future.

Most importantly,
in the present study we have suggested a simple lattice model
to describe a quasiphase transition of orientational order of water dipoles in a single file chain discussed in Ref.~\cite{Ma2017},
which accounts for
i) short-range (hydrogen bonding) and long-range (dipole-dipole) interactions
and
ii) rotations within the restricted geometry of the CNT.
The lattice model reproduces the emergence of highly ordered structure suggested in Ref.~\cite{Ma2017}
which persists in a wide range of temperatures:
The states with dipole moments oriented along the CNT axis dominate partition function at the intermediate temperatures
as evidenced by analysis of the finite-$N$ results
for the partition function, $\mu_{\parallel}$, $\vert\mu_{\perp}\vert$, the specific heat or dipole correlators.
Such a collective behavior is quite robust even with variations of the chosen parameters.
The obtained predictions are in a reasonable agreement with MD simulations reported in Sec.~\ref{sec22} and Ref.~\cite{Ma2017}.

Within the frames of the lattice model,
the hydrogen-bonded chains of the length 2 and the water molecules in the state 2 dominate in a certain temperature range $T_1\ldots T_2$
resulting in emergence of a temperature-driven orientational ordering of water molecules in CNT.
The lattice model provides estimates for $T_1$ and $T_2$, see Eq.~(\ref{13})
(in MD simulations these temperatures were roughly estimated from orientational probability distribution,
see Fig.~4 of Ref.~\cite{Ma2017}).
Moreover,
it yields new predictions for the temperature dependence of the specific heat, average length and thermal expansion or dipole correlators.
Thus, with the introduced lattice model, we have provided a new statistical mechanics perspective for understanding the behavior of the water chain inside CNT,
in particular for the emergence of three different regimes (ordered-ordered-disordered) with the temperature change.

It is worth making several general comments on the lattice model used to describe the quasiphases.
Clearly, we face a finite number of sites one-dimensional lattice model and any true phase transitions cannot be expected.
However, a gradual replacement of one quasiphase by another is possible.
The lattice model introduced in Sec.~\ref{sec3} has a number of features not typical for standard lattice models used for description of phase transitions.
First, there are $\approx 2.52$ states per site that is a consequence of the imposed restrictions.
Second,
in addition to intersite short-range and long-range interactions the model accounts for rotations at each site.
Rotations introduce some reweighting of configurations determined by interactions.
Third, the lattice model changes its volume depending on the state of lattice sites.
The reported analysis of the lattice model is based on a straightforward analytical calculations of all quantities of interest
and, therefore, is restricted to the number of sites $N=12$
(for some quantities up to $N=16$).
It might be interesting to elaborate other approaches of statistical mechanics
to examine longer chains in order to understand how characteristic features of the model evolve as $N$ increases.
Note, however, that in experiments with the water molecules in CNT,
an essentially finite-$N$ case is plausible,
when one faces many isolated water-filled CNTs of various (basically not very long) lengths with some distribution
implying a tiny probability for the formation of very short and very long chains.
Then, the dependence on $N$ vanishes after averaging by over $N$
(rather than after sending $N$ to infinity).
Other properties of the lattice model (e.g., dielectric properties) are of interest, too.

\section*{Supplementary Material}

See Supplementary Material for more quantum chemistry calculations and some properties of longer lattice chains.

\section*{Acknowledgments}

The authors thank Taras Verkholyak for discussions.
The authors thank the reviewers for constructive criticism.
The molecular dynamics calculations were performed on clusters of Ukrainian Academic Grid.
S.~M.~de Souza and O.~Rojas thank the Brazilian agencies FAPEMIG and CNPq for their partial financial support.
O.~Derzhko was supported by the Brazilian agency FAPEMIG (CEX - BPV-00090-17);
he appreciates the kind hospitality of the Federal University of Lavras in October-December of 2017.
O.~Derzhko acknowledges the kind hospitality of the ICTP, Trieste
at the activity Strongly Correlated Matter: from Quantum Criticality to Flat Bands (August 22 -- September 2, 2022)
when finalizing this paper.

\section*{Author declarations}

{\bf Conflict of interest:}

The authors have no conflicts to disclose.

{\bf Author contributions:}

O.~D. conceived the study;
M.~D. performed molecular dynamics simulations;
V.~K. performed quantum chemical calculations;
T.~K. and O.~R. performed calculations for the lattice model;
T.~C.~B., S.~M.~S., and O.~R. analyzed the data.
All authors discussed the results and commented on the manuscript.

\section*{Data availability}

The data that support the findings of this study are available within the article.

\medskip

\appendix
\section*{Supplementary Material}
\renewcommand{\theequation}{S.\arabic{equation}}
\setcounter{equation}{0}

\subsection{Quantum chemistry calculations by some other methods}

\begin{table}
\caption{Quantum chemistry predictions for the water molecule inside the CNT,
see the main text.
Semi-empirical methods AM1, PM3, and PM6.}
\vspace{2mm}
\begin{center}
\begin{tabular}{|c|c|c|c|}
\hline
                   & AM1       & PM3 & PM6
\tabularnewline
\hline
$q_{\rm O}$ ($e$)  & $-0.4348$ & $-0.4251$ & $-0.6610$
\\
\hline
$q_{\rm H}$ ($e$)  & $0.2174$ & $0.2093$ & $0.3295$
\tabularnewline
\hline
$\alpha_{\rm H-O-H}$ ($^{\circ}$)  & $103.8$ & $107.9$ & $107.89$
\tabularnewline
\hline
$r_{\rm O-O}$ (\AA)  & $2.56$ & $2.61$ & $2.605$
\tabularnewline
\hline
$r_{\rm O-H}$ (\AA) & $0.96$ & $0.96$ & $0.96$
\tabularnewline
\hline
\end{tabular}
\end{center}
\label{tab1}
\end{table}

\begin{table}
\caption{Quantum chemistry predictions for the water molecule restricted to one dimension (without CNT),
see the main text.
Semi-empirical methods AM1, PM3, and PM6.}
\vspace{2mm}
\begin{center}
\begin{tabular}{|c|c|c|c|}
\hline
                  & AM1     & PM3   & PM6
\\
\hline
$q_{\rm O}$ ($e$) & $-0.4420$ & $-0.4236$ & $-0.6789$
\\
\hline
$q_{\rm H}$ ($e$) &  0.2210 & 0.2122 & 0.3394
\tabularnewline
\hline
$\alpha_{\rm H-O-H}$ ($^{\circ}$) &  104.1 & 107.3& 104.6
\tabularnewline
\hline
$r_{\rm O-O}$ (\AA) &  2.57 & 2.6 & 2.42
\tabularnewline
\hline
$r_{\rm O-H}$ (\AA) & 0.96 & 0.96 & 0.97
\tabularnewline
\hline
\end{tabular}
\end{center}
\label{tab2}
\end{table}

\begin{table}
\caption{Quantum chemistry predictions for the water molecule restricted to one dimension (without CNT),
see the main text.
Hartree-Fock method results with the STO-2G, 6-31G, and Huzinaga MINI basis sets.
A hydrogen-bonded chain implies,
first,
a forming bond hydrogen charge $q_{\rm H}$ (the first number before slash in the third row)
and
a dangling hydrogen charge $q_{\rm H}$ (the second number after slash in the third row)
and,
second,
a shorter and a longer ${\rm O}$-${\rm H}$ bonds with $r_{\rm O-H}$ given by the first and the third numbers in the last row, respectively,
as well as a dangling hydrogen with $r_{\rm O-H}$ given by the second number in the last row.
The partial charges $q_{\rm O}$ and $q_{\rm H}$ are determined
from the Mulliken population analysis (the upper rows in the second and third rows)
or
from the L\"{o}wdin population analysis (the lower rows in the second and third rows).}
\vspace{2mm}
\begin{center}
\begin{tabular}{|c|c|c|c|}
\hline
                   & STO-2G       & 6-31G  &  MINI
\tabularnewline
\hline
$q_{\rm O}$ ($e$)  & $\begin{array}{c}
-0.2832\\-0.1930\end{array}$ & $\begin{array}{c}-0.9004\\-0.6170\end{array}$ & $\begin{array}{c}-0.6215\\-0.4650\end{array}$
\\
\hline
$q_{\rm H}$ ($e$) & $\begin{array}{c}0.1827/0.1009\\0.1304/0.0631\end{array}$ & $\begin{array}{c}0.4954/0.4055\\0.3198/0.2973\end{array}$ & $\begin{array}{c}0.3435/0.2784\\0.2583/0.2072\end{array}$
\tabularnewline
\hline
$\alpha_{\rm H-O-H}$ ($^{\circ}$)  & 99.2 & 108.7 & 103.4
\tabularnewline
\hline
$r_{\rm O-O}$ (\AA)  & 2.55 & 2.72 & 2.72
\tabularnewline
\hline
$r_{\rm O-H}$ (\AA)  & 0.99/1.00/1.61  & 0.95/0.96/1.88  & 0.99/1.00/1.82
\tabularnewline
\hline
\end{tabular}
\end{center}
\label{tab3}
\end{table}

\begin{table}
\caption{Quantum chemistry predictions for the water molecule restricted to one dimension (without CNT),
see the main text.
Density-functional-theory method (B3LYP) results,
see explanations in the title of Table~\ref{tab3}.}
\vspace{2mm}
\begin{center}
\begin{tabular}{|c|c|}
\hline
                    & B3LYP
\tabularnewline
\hline
$q_{\rm O}$ ($e$)   & $\begin{array}{c}-0.8324\\-0.7022\end{array}$
\\
\hline
$q_{\rm H}$ ($e$)   & $\begin{array}{c}0.4385/0.3920\\0.3550/0.3452\end{array}$
\tabularnewline
\hline
$\alpha_{\rm H-O-H}$ ($^{\circ}$)  & 103.0
\tabularnewline
\hline
$r_{\rm O-O}$ (\AA)   & 2.72
\tabularnewline
\hline
$r_{\rm O-H}$ (\AA)   & $0.97/0.98/1.83$
\tabularnewline
\hline
\end{tabular}
\end{center}
\label{tab4}
\end{table}

It is worth to discuss quantum chemistry predictions beyond the AM1 method
(see Sec.~\ref{sec21} and the second column in Table~\ref{tab1}).
To this end, we again use the GAMESS package \cite{Schmidt1993}.
First we consider the one-dimensional water molecules described above, however, without the (6,5) CNT
and perform
other semi-empirical calculations (Table~\ref{tab2}),
Hartree-Fock calculations (Table~\ref{tab3}),
and
also density-functional-theory calculations (Table~\ref{tab4}).
Then we return to the one-dimen\-sional water molecules inside the (6,5) CNT described above,
to illustrate the effect of the nanotube (Table~\ref{tab1}).
Note that we do not account for dispersion corrections \cite{Grimme2016} here;
accurate electronic structure calculations are far beyond the scope of the present study.

The results of two more semi-empirical calculations \cite{Christensen2016},
PM3 (Parametric Method 3)
and
PM6 (Parameterization Method 6),
are reported in Table~\ref{tab2}.
In the absence of the nanotube,
the values of $q_{\rm O}$ and $q_{\rm H}$ although change but not very dramatically,
cf., e.g., the second columns in Tables~\ref{tab1} and \ref{tab2}.
Furthermore,
while the AM1 and PM3 results for $q_{\rm O}$ and $q_{\rm H}$ differ only very slightly,
the PM6 predictions for the charge values are noticeably larger,
see the last column in Table~\ref{tab2}.

First-principle calculations using various basis sets,
STO-2G (2 primitive Gaussian orbitals are fitted to a single Slater-type orbital),
6-31G (one of Pople's split-valence basis sets),
and
Huzinaga's MINI,
are reported in Table~\ref{tab3}.
These calculations imply a hydrogen-bonded chain ground state resulting in,
first, two different charges $q_{\rm H}$ for the hydrogen forming the hydrogen bond and for the dangling hydrogen,
see two numbers separated by slash in the third row in Table~\ref{tab3}
and,
second, ${\rm O}-{\rm H}$ bonds of different lengths
[each oxygen neighbors to three hydrogens,
that is,
the two covalent hydrogens within the molecule (the one along the zig-zag and the dangling one)
and the third -- through the hydrogen bond],
see the last row with values of $r_{\rm O-H}$ in Table~\ref{tab3}.
We determined the partial charges $q_{\rm O}$ and $q_{\rm H}$ from both the Mulliken and L\"{o}wdin population analysis
and present these results in Table~\ref{tab3},
see the corresponding split for upper and lower data in the second and third row.
Again we observe a noticeable difference between the outcomes of the three calculation schemes.

We also use the B3LYP (Becke, 3-parameter, Lee-Yang-Parr) hybrid functional in the density-functional-theory method \cite{Brandenburg2019},
see Table~\ref{tab4}.
Interesting to note that the results for $q_{\rm O}$ and $q_{\rm H}$ are about 2 times larger than the AM1 predictions.

Finally,
we perform semi-empirical calculations for the (6,5) CNT,
using the structure with 362 carbon atoms and 20 hydrogen atoms added to saturate free carbon bonds on the edges of CNT
and studying 11 water molecules inside the CNT as explained above,
see Table~\ref{tab1}.
The presence of the nanotube results only in slight changes of some parameters
as can be seen by comparison of Table~\ref{tab2} and Table~\ref{tab1}.

Summarizing,
we may emphasize the diversity of quantum chemistry predictions
which suggests the necessity of further quantum-mechanical studies of water molecules in CNT.
X.~Ma {\it et al.} \cite{Ma2017} used obtained realistic charges for the water hydrogen and oxygen atoms
from a semi-empirical calculations at the AM1 level.
Herein above we present the outcomes of some other methods,
leaving for future studies an extensive quantum chemistry analysis of water molecules in CNT,
which is far beyond the scope of the present paper.

\subsection{Probabilities of short hydrogen-bonded chains for longer lattice models}

\begin{figure}[htb!]
\begin{center}
\includegraphics[clip=true,width=0.8\columnwidth]{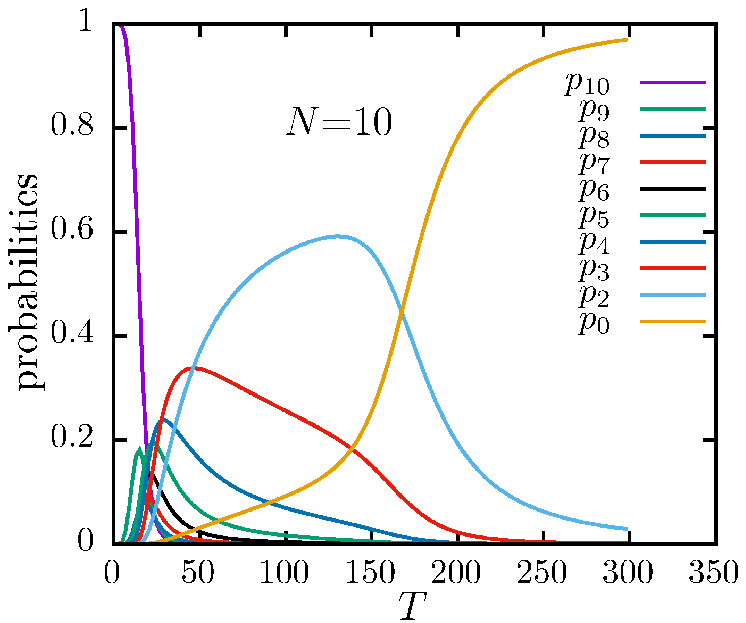}\\
\includegraphics[clip=true,width=0.8\columnwidth]{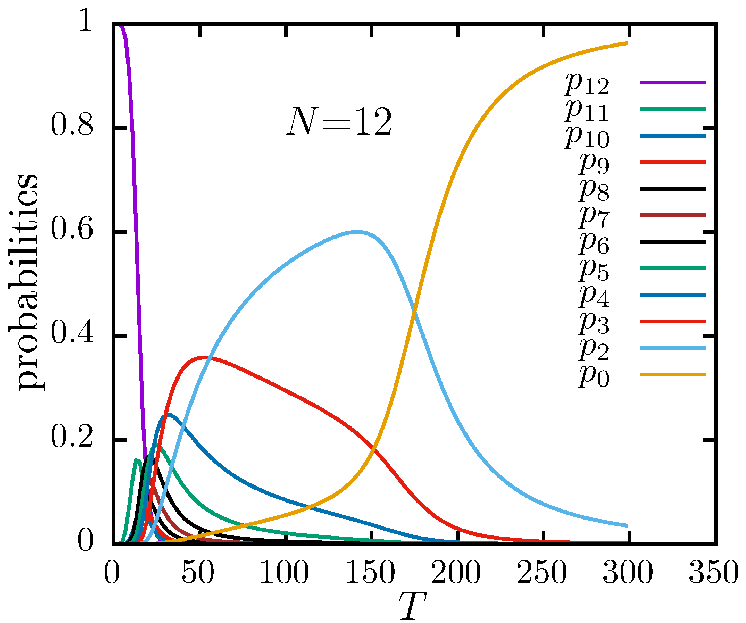}\\
\includegraphics[clip=true,width=0.8\columnwidth]{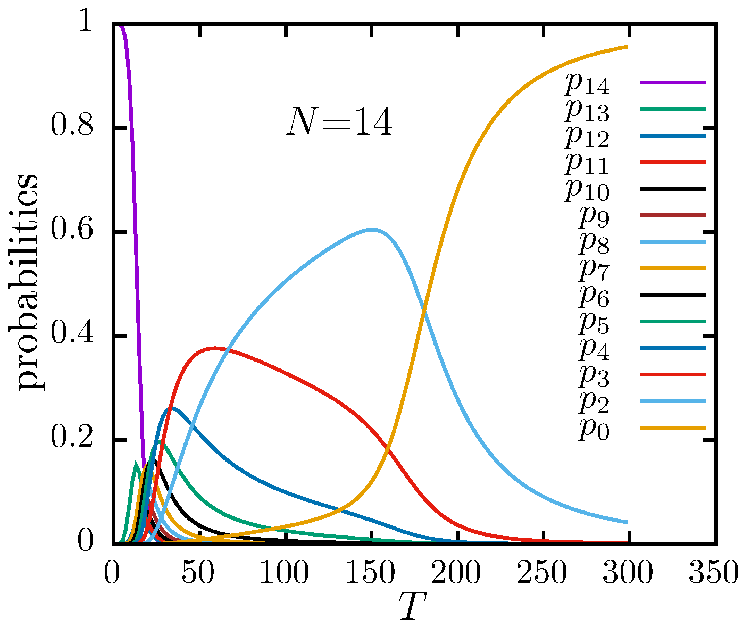}\\
\includegraphics[clip=true,width=0.8\columnwidth]{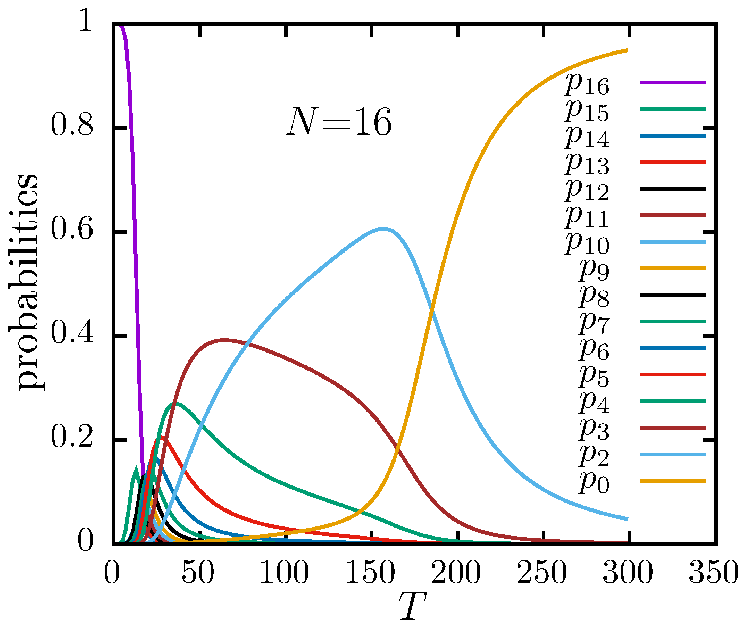}
\caption{Probabilities of various configurations versus temperature shown in Fig.~\ref{fig07}
for longer chains of $N=10,12,14,16$ sites.}
\label{fig10}
\end{center}
\end{figure}

In the main text we illustrate a role of the hydrogen-bonded chains of the length 2
reporting the temperature dependence of their contribution to thermodynamics for $N=4,6,8$ in Fig.~\ref{fig07}.
Our conclusions remain qualitatively the same for larger $N$.
This can be seen from the results reported in Fig.~\ref{fig10} which refer to the lattice model of $N=10,12,14,16$ sites.
As the computational complexity increases rapidly with the system size,
while offering no qualitative change in behavior,
considering larger $N$ looks worthless.

\end{document}